\documentclass[pre,twocolumn]{revtex4}
\usepackage[english]{babel}
\usepackage{amsmath}
\usepackage{amssymb}
\usepackage{epsfig}

\begin{document}

\title{Discrete vortices on spatially nonuniform two-dimensional electric networks}
\author{Victor P. Ruban}
\email{ruban@itp.ac.ru}
\affiliation{Landau Institute for Theoretical Physics, RAS,
Chernogolovka, Moscow region, 142432 Russia} 
\date{\today}

\begin{abstract}
Two-dimensional arrays of nonlinear electric oscillators are considered theoretically, 
where nearest neighbors are coupled by relatively small, constant, but non-equal capacitors. 
The dynamics is approximately reduced to a weakly dissipative defocusing discrete nonlinear 
Schr\"odinger equation with translationally non-invariant linear dispersive coefficients. 
Behavior of quantized discrete vortices in such systems is shown to
depend strongly on the spatial profile of the inter-node coupling as well as
on the ratio between time-increasing healing length and lattice spacing.
In particular, vortex clusters can be stably trapped for some initial period of time 
by a circular barrier in the coupling profile, but then, due to gradual dissipative 
broadening of vortex cores, they lose stability and suddenly start to move.
\end{abstract}

\maketitle
%%%%%%%%%%%%%%%%%%%%%%%%%%%%%%%%%%%%%%%%%%%%%%%%%%%%%%%%%%%%%%%%%%%%%%%%%%%%

\section{Introduction}

Nonlinear complex wave fields are known to support quantized vortices in two and three
spatial dimensions \cite{Pismen,PS1,PS2,KFC2015,SF2000,FS2001,F2009,PGFK2004}. 
Vortices have also been studied in discrete systems (on lattices; see, e.g.,
\cite{MK2001,KMFC2004,CKMF2005,KFCMB2005,LSCASS2008,KMT2011,CJKL2009,LSK2014} 
and citations therein).
As far as weakly dissipative lattice dynamics is considered, among the most popular
mathematical models are modifications of a discrete nonlinear Schr\"odinger equation (DNLSE) 
\cite{CJKL2009,LSCASS2008,KMT2011,LSK2014,HT1999,KP1992,MK2001,DKMF2003,KMFC2004,
CKMF2005,KFCMB2005,XZL2008,ACMSSW2010,RVSDK2012,ACM2014,AC2019,R2019}. 
They arise in various scientific contexts 
(but mostly in nonlinear optics \cite{LSCASS2008,KMT2011} and
in physics of nonlinear metamaterials \cite{LSK2014}), 
where we have nearly identical oscillators with their normal complex variables 
$a_n(t)=A_n(t)\exp(-i\omega_0 t)$, 
and with nonlinear frequency shifts $g|A_n|^2\ll \omega_0$.
The simplest form of DNLSE is
\begin{equation}
i(\dot A_n+\gamma \omega_0 A_n) = g|A_n|^2A_n -\frac{1}{2}\sum_{n'}c_{n,n'}A_{n'},
\label{A_n_eq}
\end{equation}
where overdot means time derivative.
A linear damping rate $\gamma\omega_0$ takes into account dissipative effects,
with small $\gamma=1/Q\ll 1$ being an inverse quality factor.
Oscillators are weakly coupled by (real) coefficients $c_{n,n'}=c_{n',n}\ll \omega_0$
(if coupling strength and/or nonlinearity level are not weak, then more 
complicated forms of DNLSE appear, including non-linearities in coupling terms
\cite{LSCASS2008,KMT2011}). 
In many interesting cases, multi-index $n$ is a node ${\bf n}=(n_1,\dots,n_d)$ 
of a simple regular lattice in one, two, or three spatial dimensions 
($d=1$, $d=2$, and $d=3$, respectively). Besides electromagnetic artificially created 
structures \cite{LSK2014}, DNLSE has been successfully applied in nonlinear optics
where it describes stationary regime of light propagation in waveguide arrays
\cite{LSCASS2008} (one-dimensional (1D) and two-dimensional (2D) cases, with time 
variable $t$ replaced by propagation coordinate $z$). 

The coupling coefficients $c_{{\bf n},{\bf n}'}$ are often considered as 
translationally invariant on the lattice and taking place between a few near neighbors. 
If they have a definite sign, then in the long-scale quasi-continuous limit we have
either defocusing regime (when $gc>0$), or focusing one (when $gc<0$).
Accordingly, different nonlinear coherent wave structures can take place in each case.
In particular, in the most well-studied focusing regime, there are highly localized 
discrete solitons and discrete vortex solitons 
(see \cite{MK2001,KMFC2004,CKMF2005,KFCMB2005,LSCASS2008}, and references therein). 
In the defocusing regime, there are dark solitons, and besides that, discrete 
analogues of superfluid quantized vortices can be excited and interact with each 
other over long distances. 

In this work, we consider discrete vortices, but in somewhat more complicated 
arrangements where coupling coefficients are not translationally invariant, 
$c_{{\bf n}+{\bf l},{\bf n}'+{\bf l}}\neq c_{{\bf n},{\bf n}'}$,
and the corresponding terms contain differences $(A_n-A_{n'})$ instead of $(-A_{n'})$,
\begin{equation}
i(\dot A_n+\gamma\omega_0 A_n) = g|A_n|^2A_n +\frac{1}{2}\sum_{n'}c_{n,n'}(A_n-A_{n'}).
\label{A_n_eq_diff}
\end{equation}
In general, equations (\ref{A_n_eq}) and (\ref{A_n_eq_diff}) are not equivalent.
Exception is for infinite and uniform lattices, where they are related to each other 
by a simple gauge transformation.

It is important that Eq.(\ref{A_n_eq_diff}), with any coefficients $c_{n,n'}$, 
admits a class of spatially uniform solutions, 
\begin{equation}
A_n=A_0\exp\big[-\gamma\omega_0 t
-ig|A_0|^2(1-e^{-2\gamma\omega_0 t})/(2\gamma\omega_0)\big]. 
\label{background}
\end{equation}
However, spatial nonuniformity of couplings should strongly affect vortex dynamics
on the above background, since vortices are known to have highly de-localized phase 
gradients even if the amplitude variation (vortex core) is localized. 
Continuous quantized vortices on spatially nonuniform backgrounds 
have been extensively studied in application to trapped Bose-Einstein condensates, 
where nonuniformity is introduced by external potential (see, e.g., Refs.
\cite{SF2000,FS2001,F2009,PGFK2004,RP1994,AR2001,GP2001,R2001,A2002,AR2002,RBD2002,AD2003,
AD2004,D2005,Kelvin_waves,ring_instability,v-2015,BWTCFCK2015,R2017-1,R2017-2,R2017-3,
R2018-1,R2018-5,reconn-2017,top-2017,WBTCFCK2017,TWK2018,TRK2019}, and citations therein).
Effects of dispersive nonuniformity are still waiting for studying. Therefore
the first goal of this work is to investigate such effects  for vortices on discrete 
lattices within model (\ref{A_n_eq_diff}). For simplicity, we consider below a square 
lattice and interactions between the nearest neighbors in the form
\begin{equation}
c_{{\bf n},{\bf n}'}=f(h[{\bf n}+{\bf n}']/2),
\label{f}
\end{equation}
where $h\ll 1$ is a lattice spacing, and $f(x,y)$ is a sign-definite function varying on 
scales $(\Delta x;\Delta y)\sim 1$. 

\begin{figure}
\begin{center}
\epsfig{file=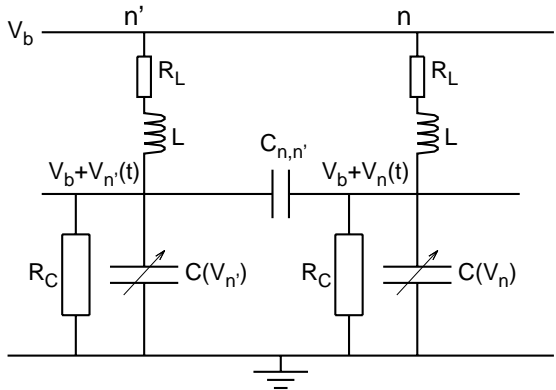, width=80mm}
\end{center}
\caption{Schematic representation of coupled oscillators. Only a fragment of whole
network is shown (two cells and coupling between them).}
\label{scheme} 
\end{figure}

Eq.(\ref{A_n_eq_diff}) with nonuniform couplings has been introduced recently in a 
formal manner as a three-dimensional (3D) discrete system supporting long-lived 
vortex knots \cite{R2019}. But no physical prototype was indicated there.
In the present work, as a possible physical implementation approximately corresponding 
to this equation, we theoretically suggest and analyze a specially designed electric
circuit network. Implementation of discrete nonlinear dynamic systems in the form of 
1D and 2D electric networks has a long and rich history 
\cite{HS1973,HCS1996,SO1999,CGBFR1998,KMR1988,MBR1994,MBR1995,LM1999,MREV2001,YMB2003,
EPSKB2010,PECCK2011,EPSCCK2013,SKVFDLR2017,PECLCK2019},
including even experimental simulations of the integrable Toda chain
\cite{HS1973,HCS1996,SO1999,CGBFR1998}. 
Major attention has been devoted to modulationally unstable systems. 
Here we consider a network possessing stable solutions (\ref{background}). 
We adopt a scheme consisting of nonlinear oscillator circuits coupled by 
relatively small, non-equal capacitors, as shown in Fig.\ref{scheme}.
It will be derived below that nonlinear constant $g$ and coupling coefficients  
$c_{{\bf n},{\bf n}'}$ appear both negative in this case, so the corresponding DNLSE 
is defocusing and appropriate for vortices. If instead of small capacitances, 
oscillators are coupled by large inductances, then a focusing DNLSE arises. 
That case has been already studied previously (on uniform lattices) in the context of 
discrete solitons, breathers and vortex solitons 
\cite{MBR1994,MBR1995,EPSKB2010,PECCK2011,EPSCCK2013}.
From a formal viewpoint, each inductor represents a separate degree of freedom.
Therefore our scheme is mathematically different. From a practical viewpoint,
small capacitors on links are more convenient than large inductors.

It should be noted that electric networks can be of macroscopic sizes and assembled 
of standard radiotechnical elements. Typical dispersive and nonlinear times can be 
about milliseconds, with the carrier frequency $\omega_0/2\pi$ of order 1 MHz.
Additional convenience of electric implementation is in easy setting the model parameters
and in controllability including arbitrary variation of coupling capacitances with time.
Moreover, flexible wires make possible to construct topologically nontrivial 
2D discrete manifolds as M\"obius strip, torus, Klein bottle, projective plane, and so on. 
This fact opens wide new perspectives in studying vortices on such discretized surfaces. 
Another important thing is that our electric scheme can be equally suitable for 
construction of 3D nonlinear lattices. 
A practical problem is only in a very big number of elements.
So, to observe interesting nonlinear behavior of vortices, in a 2D lattice we need about 
${\mathsf N}_{\rm 2D}\sim 10^3-10^4$ individual oscillators, while for a 3D lattice 
the required number is ${\mathsf N}_{\rm 3D}\sim 10^5-10^6$. Therefore planar 
constructions seem more realistic at the moment. 

Since the electric  model is very promising, we put also the second goal in this work, 
that is to simulate the dynamics directly within equations of motion governing the scheme 
in Fig.1, and then compare the results with DNLSE simulations.  

This article is organized as follows. In section II we introduce the theoretical model
and derive the corresponding DNLSE together with the parameters. Some technical details
about DNLSE are included because it is more easy for theoretical analysis than the basic
system of circuit equations. In section III, we generally analyze  vortex motion in 
2D case, with orientation on quasi-continuous limit. Special attention is given there to
coupling profiles with a barrier. This feature is new in comparison with Ref.\cite{R2019}.
In section IV, we present some numerical results demonstrating nontrivial behavior 
of interacting vortices in discrete, spatially nonuniform, weakly dissipative
2D systems. Both DNLSE and the original system of circuit equations are simulated.
In particular, it will be shown that depending on parameters,
vortex clusters can be stably trapped for some initial period of time by a 
circular barrier in function $f$ profile, but then, due to gradual dissipative 
broadening of vortex cores, they lose stability and suddenly start to move 
in a complicated manner, some of vortices penetrating the barrier.
Finally, section V contains a brief summary of the work.

\section{Model description and basic equations}

In the beginning, we describe our simple scheme (see Fig.\ref{scheme}).
Let each electric oscillator in the network consist of a coil with inductance $L$ 
and small active resistance $R_L$, connected in series to a voltage-dependent 
differential capacitance (varicap) $C_v(V)=dq/dV$, where $q$ is electric charge.
A reverse-biased varactor diode is implied or another nonlinear capacitor
(perhaps in parallel with an ordinary capacitor).
The varicap is characterized by a large shunt resistance $R_C$ (for leakage current).
For simplicity, we assume $R_C=const$ thus neglecting nonlinearity in dissipation.
The remaining end of the coil is connected to a d.c. bias voltage $V_b$, 
while the remaining contact of the varicap is grounded. A voltage at the contact 
between the coil and the varicap is $V_b+V_n(t)$. Functional dependencies 
$C(V_n)=C_v(V_b+V_n)$ differ for devices fabricated under different technologies, 
so many expressions were suggested to approximate them.
In particular, for a reverse-biased diode in parallel with a constant capacitor,
the following combined formula is able to ensure good accuracy within a sufficiently 
wide voltage range (see, e.g., \cite{MBR1995,HCS1996,CGBFR1998,EPSKB2010,PECCK2011}), 
\begin{equation}
C(V_n)=C_0\Big[\mu+\frac{(1-\mu)}{(1+V_n/V_*)^\nu}+\eta e^{-\kappa V_n}\Big]/(1+\eta),
\label{C_V}
\end{equation}
with fitting parameters $C_0=C(0)$, $V_*$, $\mu$, $\nu$, $\eta$, and $\kappa$.
Here $0<\mu<1$ takes into account an ordinary capacitor in parallel, while
$0.3\lesssim \nu\lesssim 6.0$ is related to the diode 
(by the way, for the Toda lattice implementation, one needs to take diodes with $\nu=1$).
Very often in theoretical studies it is put $\eta=0$.
In some research works, a different kind of variable capacitor is also considered, 
with $C(V_n)=C_0(1+V_n^2/V_*^2)$ \cite{SKVFDLR2017}. 
Such a symmetric dependence is possible in devices using 
special nonlinear dielectric films \cite{MFSY2016}.
In any case, (additional) accumulated electrostatic energy 
at the varicap is given by formula
\begin{equation}
W(V_n)=\int_{0}^{V_n}C(u)u du,
\end{equation}
while a.c. electric charge is
\begin{equation}
q_n=q(V_n)=\int_{0}^{V_n}C(u) du.
\end{equation}
Taking the inverse relation, we have $V_n=U(q_n)$. Equation of motion for a single 
oscillator circuit, with dissipative terms neglected, is $L\ddot q_n+U(q_n)=0$. 
It will be important for our purposes that a nonlinear frequency shift 
can be negative in this dynamics. Of course, fully nonlinear regime should be studied 
numerically, but analytical investigations may be based on the expansion
\begin{equation}
U(q_n)=C_0^{-1}\big[q_n +\alpha q_n^2+\beta q_n^3+\cdots \big]
\end{equation}
assuming relatively small amplitudes. 
Then a frequency shift for weakly nonlinear regime is known to be 
\begin{equation}
\Delta\omega=\omega_0(3\beta/8-5\alpha^2/12)q_0^2, 
\end{equation}
with $\omega_0=2\pi/T_0=1/\sqrt{LC_0}$, and $q_0$ being an amplitude of the main harmonics.

The inverse quality factor of the oscillator is apparently
\begin{equation}
\gamma=\left(R_L\sqrt{C_0/L}+R_C^{-1}\sqrt{L/C_0}\right)/2.
\end{equation}
It is presumed very small (as we will see below, the most interesting 
things happening with vortices begin at $Q\gtrsim 10^4$).
For example, with $L=5.0\times 10^{-4}$ H, $C_0=5.0\times 10^{-10}$ F, $R_L<0.1$ Ohm, and
$R_C>10^7$ Ohm, we have $\omega_0=2.0\times 10^6$ rad/s, corresponding to a frequency about
0.3 MHz, and a sufficiently high quality factor $Q>10^4$. Perhaps, even smaller values 
of $R_L$ and larger values of $R_C$ can be achieved at reasonably low temperatures, making 
$Q\gtrsim 10^5$.

There are also weak ordinary capacitors $C_{n,n'}\ll C_0$ inserted between points 
$V_n$ and $V_{n'}$. They unite individual oscillators into a whole network.

Equations of motion for the united system can be derived in a very simple manner. Indeed, 
electric current through the coil is $I_n$, while currents through the capacitors 
are $C(V_n)\dot V_n$ and $C_{n,n'}(\dot V_n -\dot V_{n'})$. 
Leakage current parallel to varicap is $V_n/R_C$. Thus we obtain equations
\begin{equation}
C(V_n)\dot V_n+\sum_{n'} C_{n,n'}(\dot V_n -\dot V_{n'})+\frac{V_n}{R_C}= I_n.
\label{current}
\end{equation}
A voltage difference at the coil is $L \dot I_n+R_L I_n$. In sum with $V_b+V_n$ it should 
give $V_b$. Therefore we have the second sub-set of equations, closing the system,
\begin{equation}
L \dot I_n + V_n + R_L I_n = 0.
\label{voltage}
\end{equation}
It is clear that our system admits a class of $n$-independent solutions related to 
Eq.(\ref{background}), when each node oscillates as if there were no couplings. 

It is not so obvious at first glance but can be easily checked that without 
dissipative terms containing active resistances $R_L$ and $R_C$, 
equations (\ref{current})-(\ref{voltage}) correspond to a Lagrangian 
system with the Lagrangian function
\begin{eqnarray}
{\mathsf L}&=&\sum_n \frac{L}{2}\Big[C(V_n)\dot V_n
+\sum_{n'} C_{n,n'}(\dot V_n -\dot V_{n'})\Big]^2
\nonumber\\ 
&&-\sum_n W(V_n)-\sum_{n,n'}\frac{C_{n,n'}}{4}(V_n - V_{n'})^2.
\end{eqnarray}

Equations of motion in the form (\ref{current})-(\ref{voltage}) are suitable enough for 
numerical simulations, but difficult for theoretical analysis. Therefore our next steps
will be to rewrite the Lagrangian in terms of charges $q_n$, and then introduce
a Hamiltonian description. It is convenient to adopt non-dimensionalization
(voltage in units $V_*$, charge in units $C_0 V_*$, time in units $1/\omega_0$), 
formally corresponding to $L=1$, $C_0=1$. Then, in the first order on small quantities 
$\bar c_{n,n'}=C_{n,n'}/C_0$, and retaining only main terms on oscillation amplitudes
in the couplings, we have 
\begin{eqnarray}
{\mathsf L}&\approx&\sum_n \Big[\frac{\dot q_n^2}{2}
-\frac{q_n^2}{2}-\alpha\frac{q_n^3}{3} -\beta\frac{q_n^4}{4}\Big]\nonumber\\
&+&\frac{1}{4}\sum_{n,n'}\bar c_{n,n'}[2(\dot q_n -\dot q_{n'})^2-(q_n -q_{n'})^2].
\end{eqnarray}
The canonical momenta for this Lagrangian are
\begin{equation}
p_n=\dot q_n+2\sum_{n,n'}\bar c_{n,n'}(\dot q_n -\dot q_{n'}).
\end{equation}
Inverse relations, again with the first-order accuracy on $\bar c_{n,n'}$, are easily
obtained as
\begin{equation}
\dot q_n\approx p_n-2\sum_{n,n'}\bar c_{n,n'}(p_n -p_{n'}).
\end{equation}
As the result, the Hamiltonian function of weakly interacting oscillators acquires 
the following form,
\begin{eqnarray}
{\mathsf H}&\approx&\sum_n \Big[\frac{p_n^2}{2}
+\frac{q_n^2}{2}+\alpha\frac{q_n^3}{3} +\beta\frac{q_n^4}{4}\Big]\nonumber\\
&-&\frac{1}{4}\sum_{n,n'}\bar c_{n,n'}[2(p_n -p_{n'})^2-(q_n -q_{n'})^2].
\end{eqnarray}
When an oscillator is taken separately, then there exists a weakly nonlinear 
canonical transform,
\begin{eqnarray}
q_n&\approx&\tilde q_n -\frac{\alpha}{3}(\tilde q_n^2+2\tilde p_n^2)\nonumber\\
&+&\frac{\tilde q_n}{16}\Big[\Big(\frac{25}{9}\alpha^2\!-\!\frac{5}{2}\beta\Big)\tilde q_n^2
+\Big(\frac{13}{9}\alpha^2\!-\!\frac{9}{2}\beta\Big)\tilde p_n^2\Big],
\end{eqnarray}
\begin{eqnarray}
p_n&\approx&\tilde p_n +\frac{2\alpha}{3}\tilde p_n\tilde q_n\nonumber\\
&-&\frac{\tilde p_n}{16}\Big[\Big(\frac{11}{9}\alpha^2\!-\!\frac{15}{2}\beta\Big)\tilde q_n^2
+\Big(\frac{47}{9}\alpha^2\!-\!\frac{3}{2}\beta\Big)\tilde p_n^2\Big],
\end{eqnarray}
such that combination $a_n=(\tilde q_n +i \tilde p_n)/\sqrt{2}$ (the normal complex variable)
is related to the action-angle variables $S_n$ and $\phi_n$ by formula
$a_n=\sqrt{S_n}\exp(i\phi_n)$. That transform excludes third-order terms from the partial
Hamiltonians. Neglecting again nonlinearities in the couplings, we reduce the total 
Hamiltonian to the following expression:
\begin{eqnarray}
{\mathsf H}&\approx&\sum_n \big(|a_n|^2+\frac{g}{2}|a_n|^4\big)\nonumber\\
&-&\frac{1}{4}\sum_{n,n'}\bar c_{n,n'}(a_n -a_{n'})(a^*_n -a^*_{n'})\nonumber\\
&+&\frac{3}{8}\sum_{n,n'}\bar c_{n,n'}[(a_n -a_{n'})^2+(a^*_n -a^*_{n'})^2],
\label{H_a}
\end{eqnarray}
where the nonlinear coefficient is $g=(3\beta/4-5\alpha^2/6)$.
In terms of $a_n$, Hamiltonian equations of motion are 
$i\dot a_n=\partial {\mathsf H}/\partial a_n^*$.
In the main approximation, $a_n$ behaves proportionally to $\exp(-it)$, 
since the nonlinearity and the couplings are weak. Therefore, the last double
sum in Eq.(\ref{H_a}) contains quickly oscillating quantities which are not important 
after averaging. Introducing slow envelopes $A_n=a_n\exp(it)$ and taking into account 
linear damping (not covered by Hamiltonian theory), we arrive at Eq.(\ref{A_n_eq_diff}), 
with negative $c_{n,n'}=-\bar c_{n,n'}$. Nonlinear coefficient $g$, for physically relevant
parameters in Eq.(\ref{C_V}), appears also negative. In particular, if $\eta=0$, then
\begin{equation}
g=\frac{\nu(1-\mu)}{24}[-3+\nu(1-4\mu)]. 
\end{equation}
It is very important that a non-zero value of $\mu$, corresponding to a constant capacitor 
in parallel with the diode, results in stronger negative frequency shift.
For example, with $\nu=2$ and $\mu=0.5$, we have $g=-5/24$, while for $\nu=2$ and $\mu=0$
it is  $g=-1/12$.

\section{Analysis of vortex motion in 2D}

As far as our goal is to study vortices on 2D networks, it is convenient to introduce 
new complex variables $\psi_n(t)$ through the following substitution 
(compare to Ref.\cite{R2019} where positive frequency shift was considered):
\begin{equation}
A_n(t)=A_0\psi^*_n(t)\exp[-\gamma t-i\varphi(t)],
\end{equation}
where real $A_0$ is a typical amplitude at $t=0$, and 
$\varphi(t)=gA_0^2(1-e^{-2\gamma t})/(2\gamma)$. As the result, we reduce our dissipative
autonomous system to a non-autonomous Hamiltonian system,
\begin{equation}
i\dot \psi_n=\sum_{n'}\frac{\bar c_{n,n'}}{2}(\psi_n-\psi_{n'}) 
+|gA_0^2|e^{-2\gamma t}(|\psi_n|^2-1)\psi_n.
\label{psi_eq}
\end{equation}
Let a typical value of $\bar c_{n,n'}$ be $\bar c\ll 1$. For purposes of further analysis,
we introduce  a slow time $\tau=h^2\bar c t$ and small parameters, 
\begin{equation}
\delta=\gamma/(h^2\bar c )\ll 1,\qquad \xi=(h^2 \bar c/|gA_0^2|)^{1/2}\ll 1.
\end{equation}
Then Eq.(\ref{psi_eq}) takes the following form,
\begin{equation}
i\frac {d \psi_{\bf n}}{d\tau }=
\sum_{\bf n'}\frac{F_{{\bf n},{\bf n}'}}{2h^2}(\psi_{\bf n}-\psi_{\bf n'}) 
+\frac{e^{-2\delta \tau}}{\xi^2}(|\psi_{\bf n}|^2-1)\psi_{\bf n},
\label{psi_eq_tau}
\end{equation}
where ${\bf n}'$ are the nearest neighbors for ${\bf n}$ on square lattice, 
$F_{{\bf n},{\bf n}'}=F(h[{\bf n}+{\bf n}']/2)$, and $F({\bf r})\sim 1$ is a non-negative 
function. In the continuous limit, the above equation reduces to a defocusing NLSE 
with spatially variable dispersion coefficient and time-dependent nonlinear coefficient, 
\begin{equation}
i\psi_\tau=-\frac{1}{2}\nabla\cdot \left[F({\bf r})\nabla\psi\right] +
\frac{e^{-2\delta \tau}}{\xi^2}(|\psi|^2-1)\psi.
\label{psi_eq_tau_contin}
\end{equation}
We are interested in vortices on constant background $\psi_0=1$.
It is clear from the equation above that intervals $\Delta\tau\sim 1$ 
are typical vortex turnover times in the system, $\xi$ is a typical relative 
healing length at $\tau=0$, while 
\begin{equation}
\tilde\xi({\bf r},\tau)=\xi e^{\delta\tau}\sqrt{F({\bf r})}
\label{tilde_xi}
\end{equation}
is a local relative vortex core width. 
Vortices described by Eq.(\ref{psi_eq_tau_contin}) have been analyzed in Ref.\cite{R2019} 
for 3D case. Applying similar analysis to 2D situation, we easily obtain that coordinates 
$x_j$ and $y_j$ of $N$ ``point'' vortices are canonically conjugate quantities (up to vortex 
signs $\sigma_j=\pm 1$). On not very long times and for small $\xi$, 
when $\xi_{eff}=\xi \exp(\delta\tau)\ll 1$, vortex motion is 
approximately described by a time-dependent Hamiltonian function 
(compare to Refs.\cite{R2017-1,R2017-2}),
\begin{eqnarray}
&&H=\sum_j\sigma_j^2 {\cal E}({\bf r}_j,\tau)
+{\sum_{j, k}}'\frac{\sigma_j \sigma_k}{2}G({\bf r}_j,{\bf r}_k),
\label{H_v}\\
&&{\cal E}({\bf r},\tau)\approx 
\frac{1}{2}G({\bf r}-{\bf e}\tilde\xi({\bf r},\tau)/2,{\bf r}+{\bf e}\tilde\xi({\bf r},\tau)/2),
\end{eqnarray}
where the prime means omitting diagonal terms in the double sum determining pair 
interactions between vortices, ${\bf e}$ is a unit vector,
and a two-dimensional Green function $G({\bf r},{\bf r}_1)$ satisfies equation
\begin{equation}
-\nabla_{\bf r}\cdot\frac{1}{F({\bf r})}\nabla_{\bf r}G({\bf r},{\bf r}_1) =
2\pi\delta_{\rm Dirac}({\bf r}-{\bf r}_1).
\label{G}
\end{equation}
The physical meaning of $G({\bf r},{\bf r}_1)$ can be explained as follows. 
Let $\psi=\sqrt{\rho}\exp(i\Phi)$ be the Madelung transform, and 
${\bf J}=\rho F({\bf r})\nabla\Phi$ be a ``current density'' (in the hydrodynamic sense)
for Eq.(\ref{psi_eq_tau_contin}). In the ``long-scale'' hydrodynamic regime,
away from vortex cores we have $\rho\approx 1$ and thus $\nabla\cdot{\bf J}\approx 0$, 
so a stream function $\Theta$ exists for 2D vector field  $F({\bf r})\nabla\Phi$. 
Since $\Phi$-field created by vortices is not single-valued and has singularities, 
it satisfies equation 
$\mbox{curl}_{2D}\nabla\Phi=2\pi\sum_j \sigma_j \delta_{\rm Dirac}({\bf r}-{\bf r}_j)$.
Therefore we have a partial differential equation determining the stream function,
\begin{equation}
-\nabla_{\bf r}\cdot\frac{1}{F({\bf r})}\nabla_{\bf r}\Theta({\bf r}) =
2\pi\sum_j\sigma_j\delta_{\rm Dirac}({\bf r}-{\bf r}_j).
\end{equation}
So $G({\bf r},{\bf r}_1)$  is a stream function
created at point ${\bf r}$ by a vortex placed in point ${\bf r}_1$.
Expression (\ref{H_v}) for vortex Hamiltonian $H$ then follows from appropriately 
regularized ``kinetic energy'' integral
\begin{equation}
2\pi H=\frac{1}{2}\int \frac{(\nabla\Theta)^2}{F({\bf r})}d^2{\bf r}. 
\end{equation}

It follows from Eq.(\ref{G}) that 
\begin{equation}
 G({\bf r}_1,{\bf r}_2)= \tilde\theta({\bf r}_1,{\bf r}_2)-
 \sqrt{F({\bf r}_1)F({\bf r}_2)}\ln|{\bf r}_1-{\bf r}_2|,
\end{equation}
with some smooth function $\tilde \theta({\bf r}_1,{\bf r}_2)\sim 1$. 
Therefore the self-energy is
\begin{equation}
{\cal E}({\bf r},\tau)= \theta({\bf r})-\frac{1}{2}
F({\bf r})\left[\ln\left(\xi\sqrt{F({\bf r})}\right)+\delta \tau\right],
\end{equation}
where $\theta({\bf r})=\tilde \theta({\bf r},{\bf r})/2$.

\begin{figure}
\begin{center}
\epsfig{file=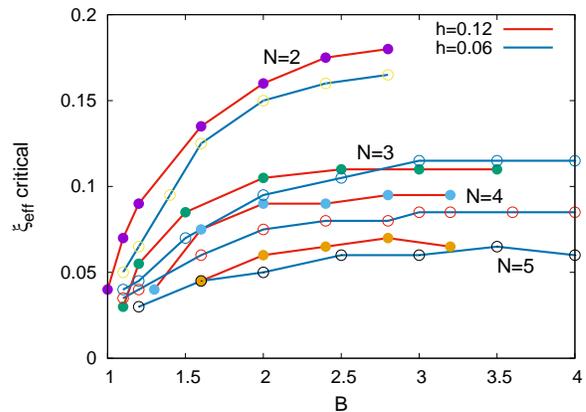, width=80mm}
\end{center}
\caption{Critical values of $\xi_{eff}$ found numerically by minimizing the 
Hamiltonian (\ref{H_xi_delta}), starting with a small $\xi_{eff}$ and increasing
it by small steps until cluster destruction.}
\label{critical_xi} 
\end{figure}

In particular, we may take circularly symmetric profile $F(r)$, with $r=\sqrt{x^2+y^2}$,
and roughly (with a logarithmic accuracy) estimate energy of a vortex cluster 
in the form of a regular $N$-polygon,
\begin{equation}
E_N(r,\tau)\approx\frac{N}{2} F(r)[\Lambda(\tau)-(N-1)\ln(r)],
\label{E_N}
\end{equation}
where $\Lambda(\tau)=[\ln(1/\xi)-\delta \tau]=-\ln(\xi_{eff})$ 
is a logarithmically large quantity. 
It is not difficult to understand that if $F(r)$ has a barrier at some finite $r_b$, 
and $N$ is not too large, then expression (\ref{E_N}) may have a minimum at some 
$0<r_*<r_b$. Thus, while $\xi_{eff}$ is less than a critical value, such a profile 
is able to trap vortex cluster. 

Discreteness (finite $h$) acts also to stabilize vortex configurations because, 
while $\xi_{eff}\lesssim h$, the lattice tends to create 
local minima (in inter-node vortex center positions) 
for the Hamiltonian corresponding to Eq.(\ref{psi_eq_tau}),
\begin{equation}
\tilde{\mathsf H}=\sum_{\bf n,n'}
\frac{F_{{\bf n},{\bf n}'}}{4h^2}|\psi_{\bf n}-\psi_{\bf n'}|^2 
+\sum_{\bf n}\frac{e^{-2\delta \tau}}{2\xi^2}(|\psi_{\bf n}|^2-1)^2.
\label{H_xi_delta}
\end{equation}
Fig.\ref{critical_xi} illustrates this fact for a particular case of ``rectangular'' barrier 
($F(r)=1$ if $r^2 < 1$, and $F(r)=B>1$ if $1\leq r^2 < 3$; otherwise $F=0$). There,
for different $N$, and for two different values of $h$,
numerical estimates are presented how critical value of parameter $\xi_{eff}$ depends
on barrier height $B$. It is seen that spatial nonuniformity of links has a strong
influence on vortex stability for $1\lesssim B\lesssim 3$. However, saturation on
larger $B$ is still waiting for explanation.

So we can expect stable trapping of a few vortices of the same sign within a domain 
surrounded by the barrier. However, as time increases, function $\Lambda(\tau)$ decreases, 
and therefore vortex configuration should suddenly become unstable at some moment.
In the next section, we numerically verify such a scenario within Eq.(\ref{psi_eq_tau}),
and then within Eqs.(\ref{current})-(\ref{voltage}).

\begin{figure}
\begin{center}
\epsfig{file=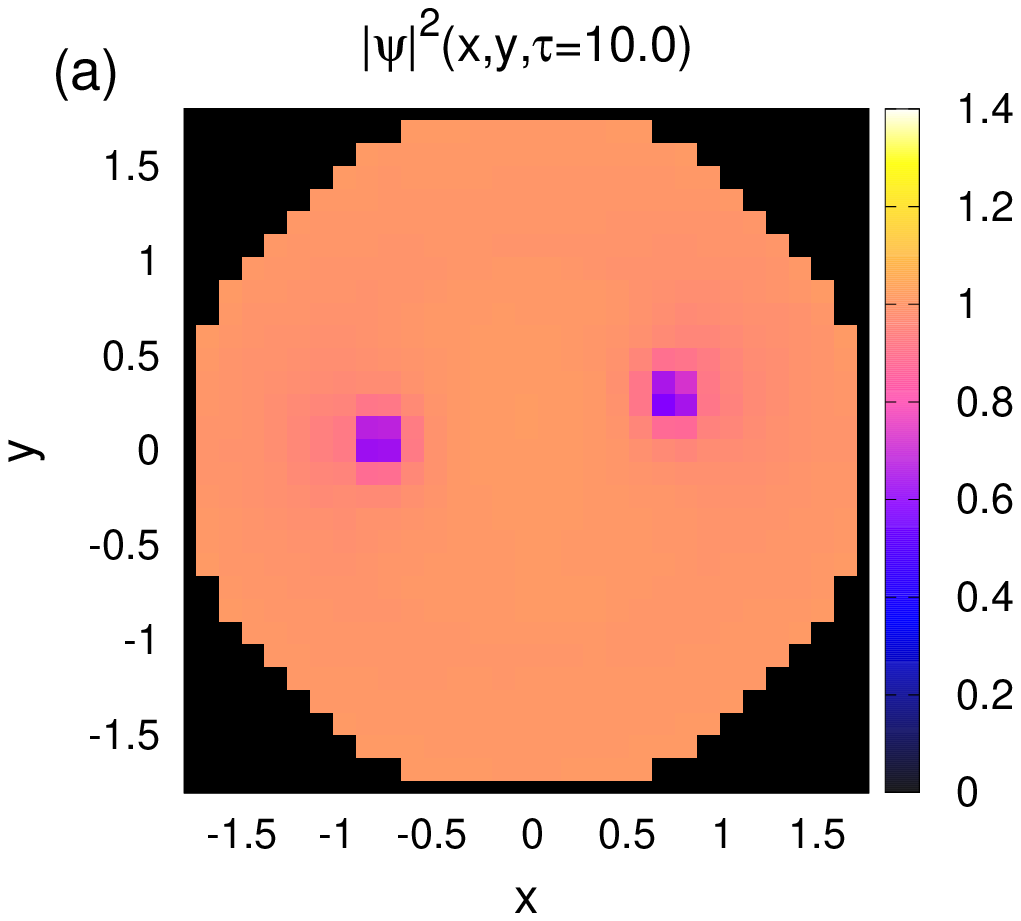, width=42mm}
\epsfig{file=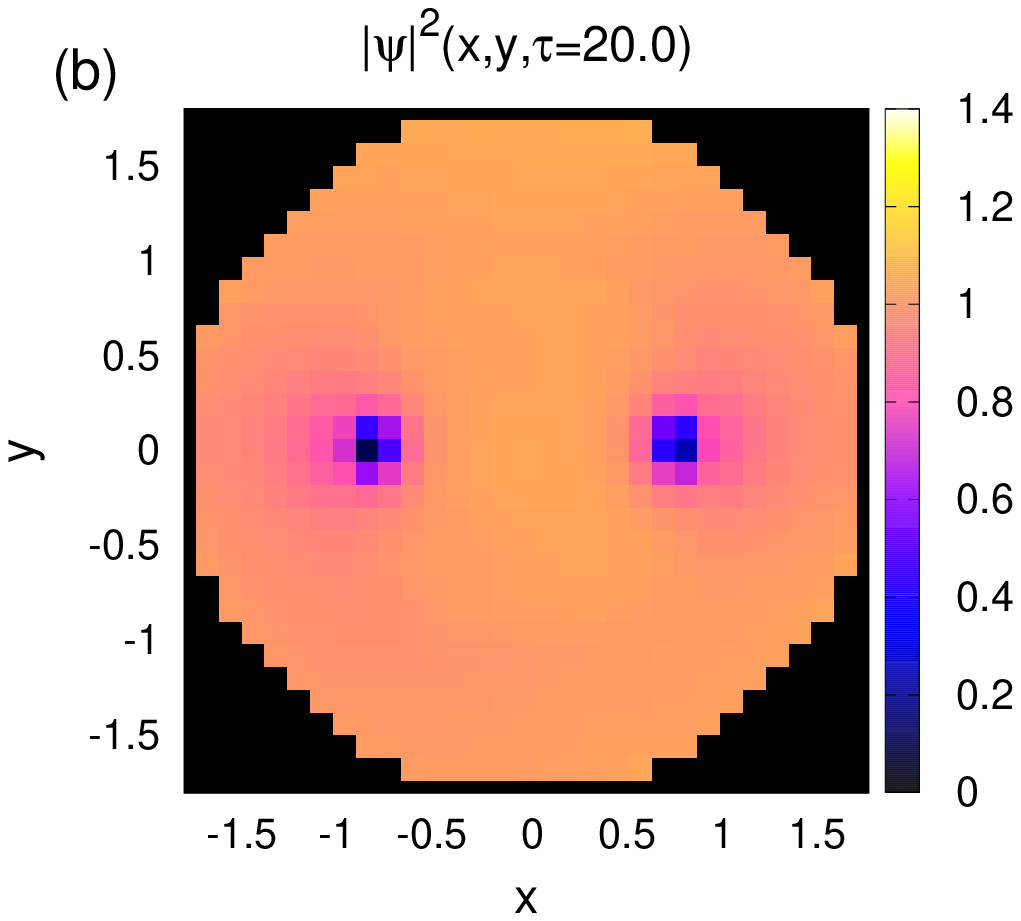, width=42mm}\\
\epsfig{file=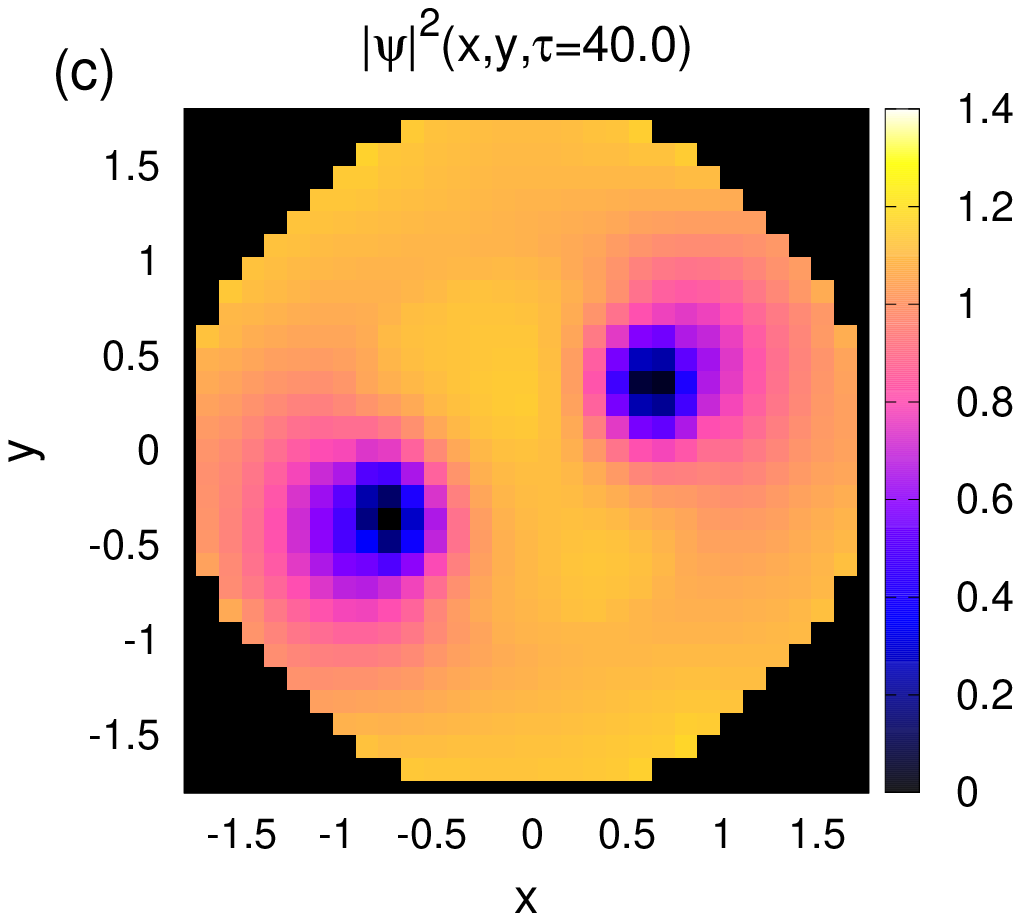, width=42mm}
\epsfig{file=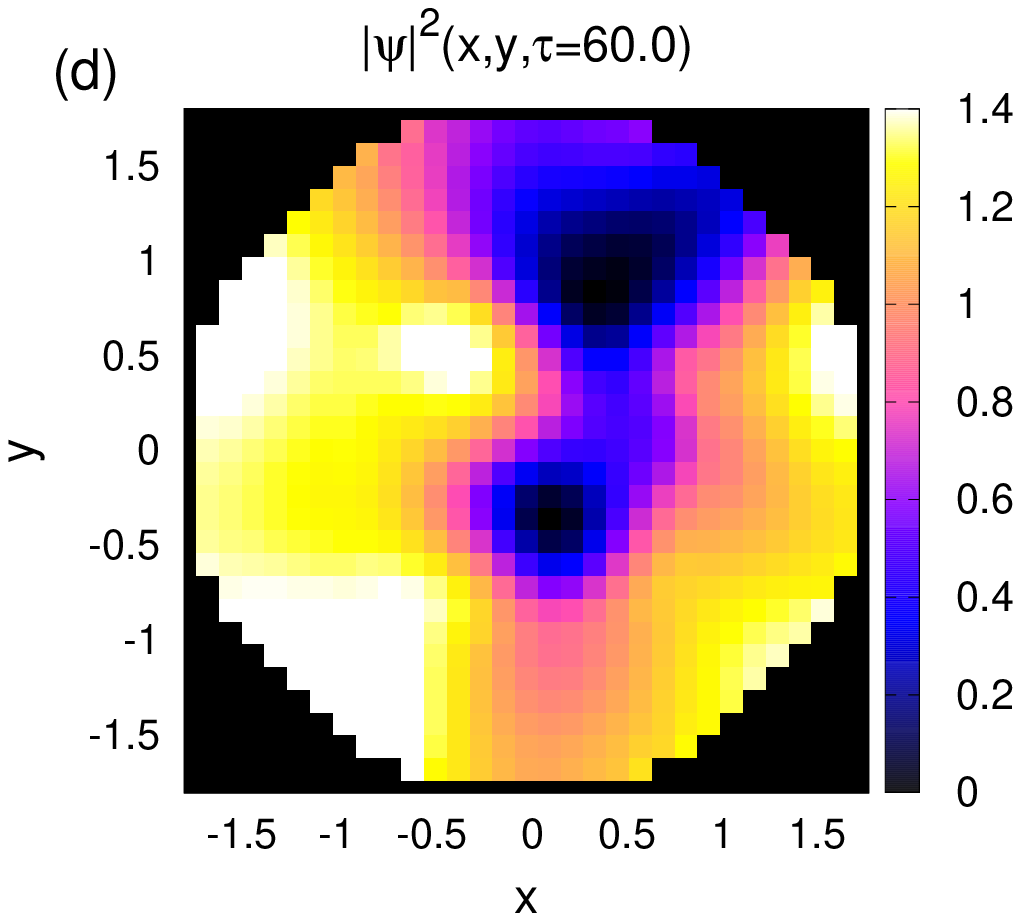, width=42mm}\\
\epsfig{file=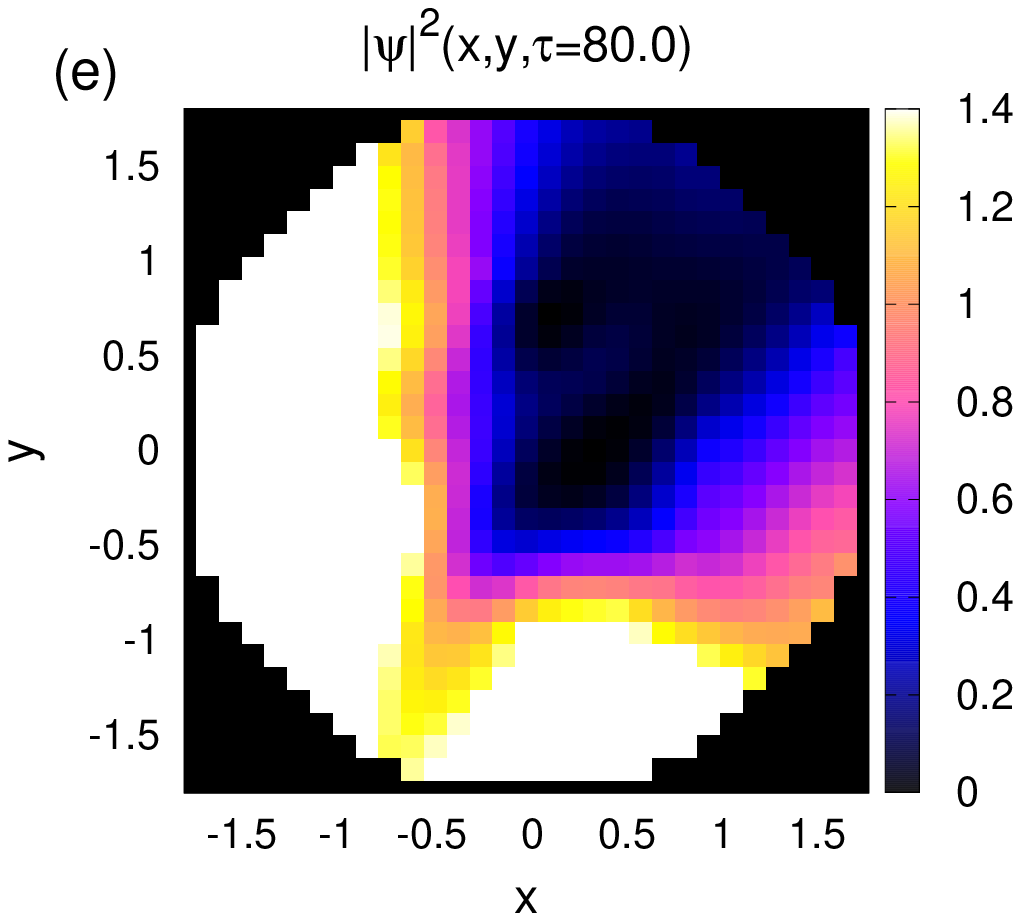, width=42mm}
\epsfig{file=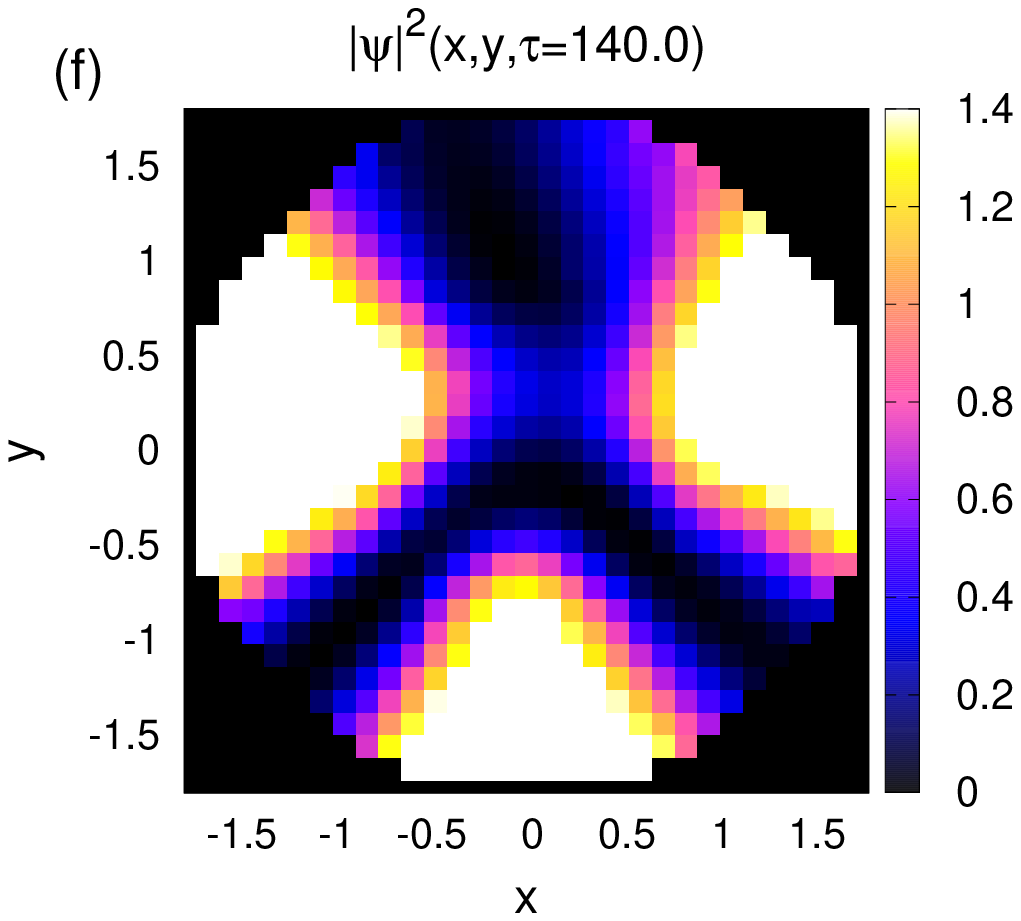, width=42mm}
\end{center}
\caption{An example of evolution of two vortices in DNLSE.}
\label{N2} 
\end{figure}
\begin{figure}
\begin{center}
\epsfig{file=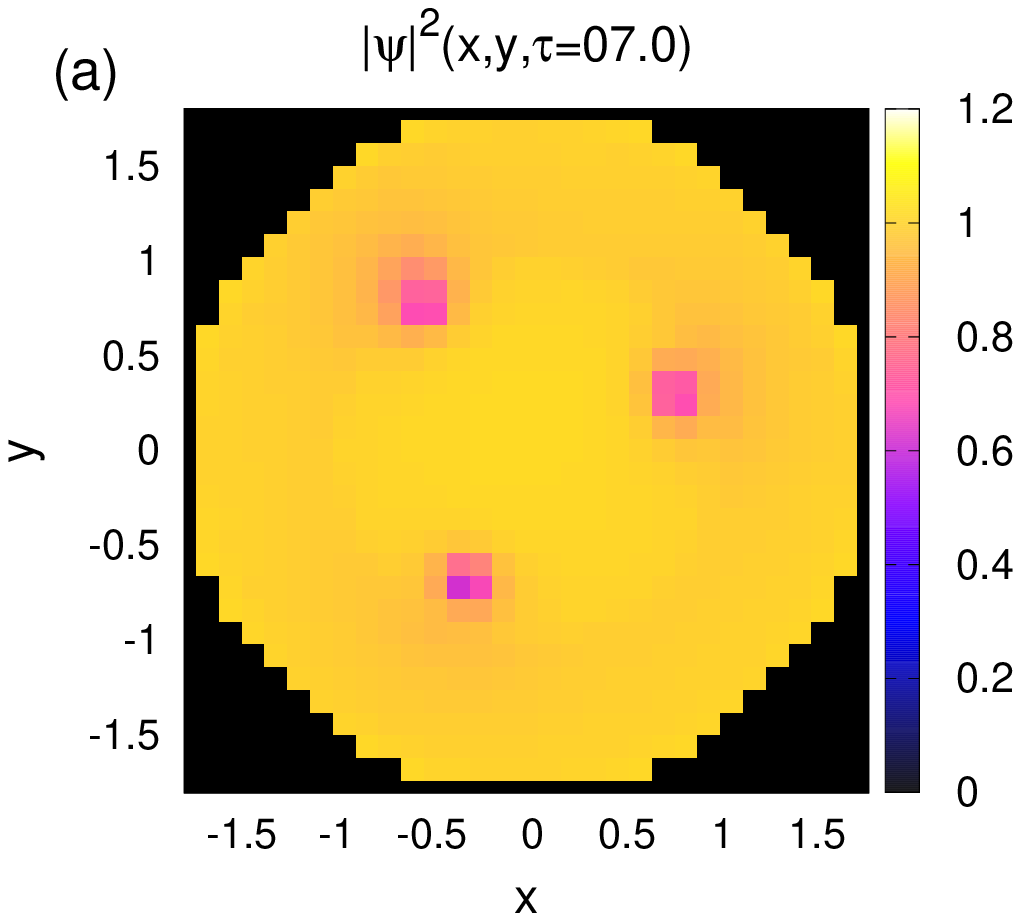, width=42mm}
\epsfig{file=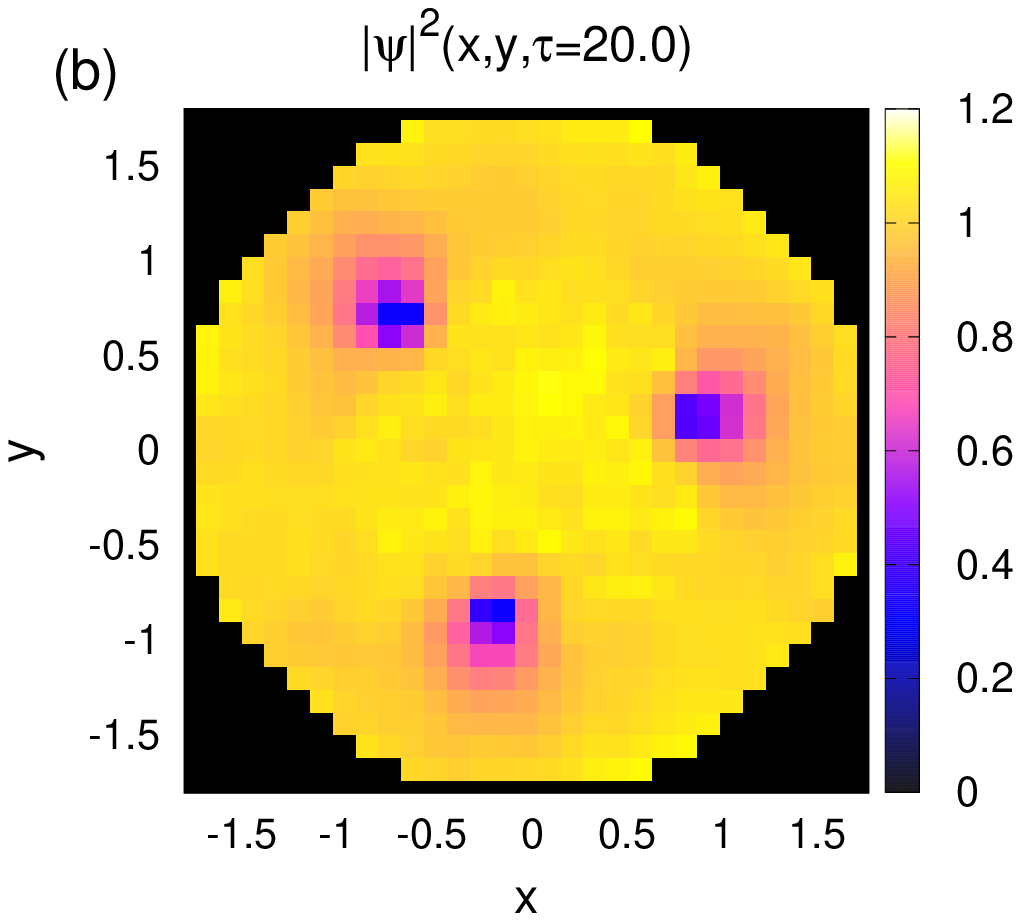, width=42mm}\\
\epsfig{file=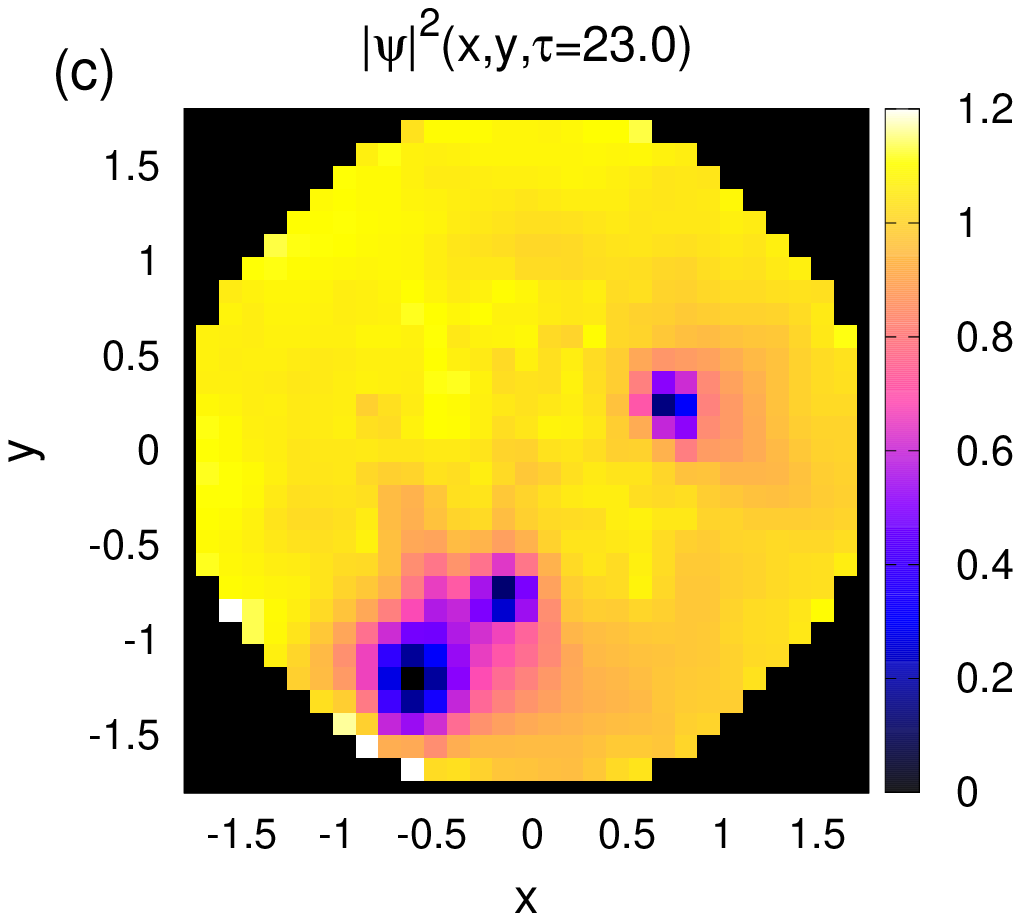, width=42mm}
\epsfig{file=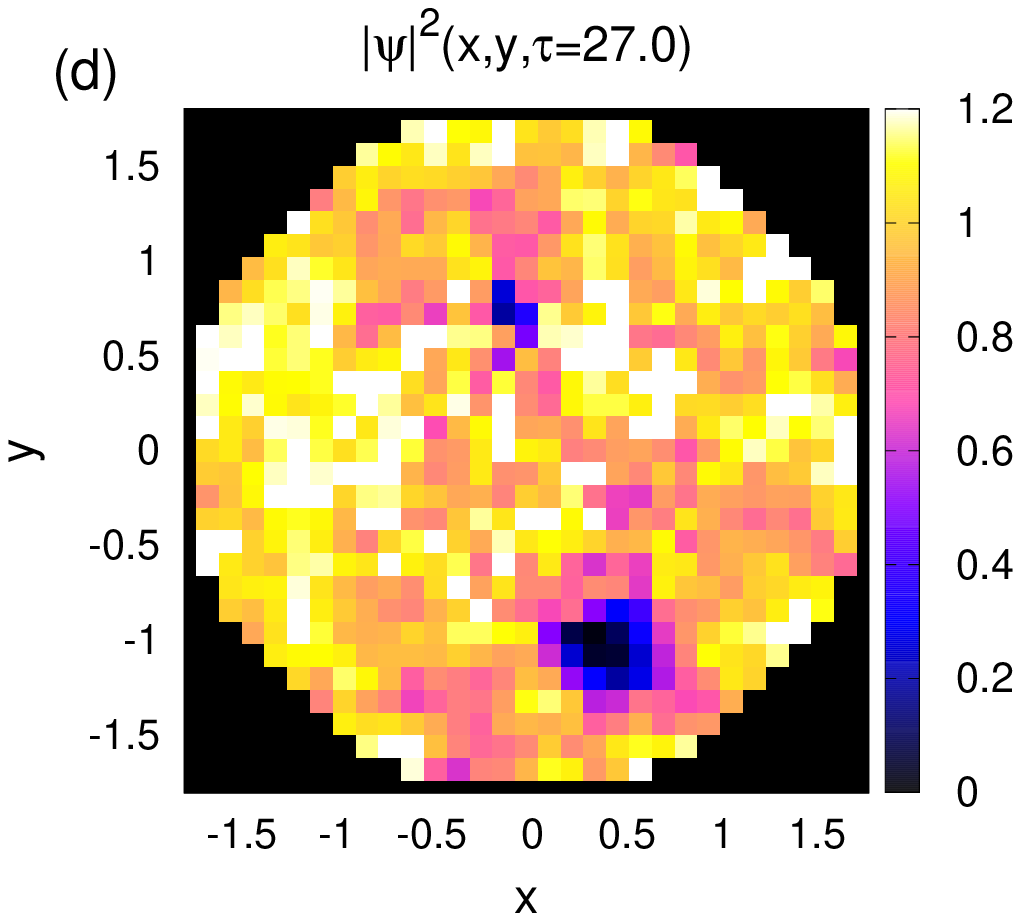, width=42mm}
\end{center}
\caption{An example of evolution of three vortices in DNLSE.}
\label{N3} 
\end{figure}
\begin{figure}
\begin{center}
\epsfig{file=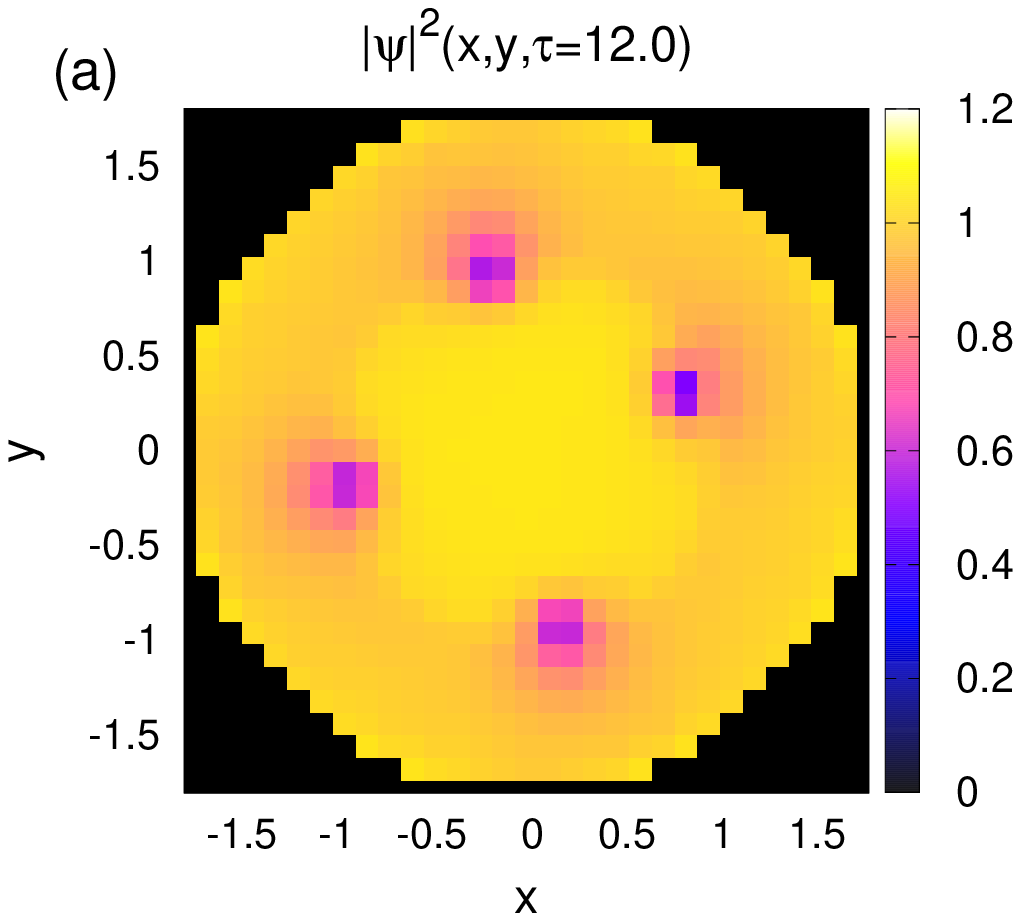, width=42mm}
\epsfig{file=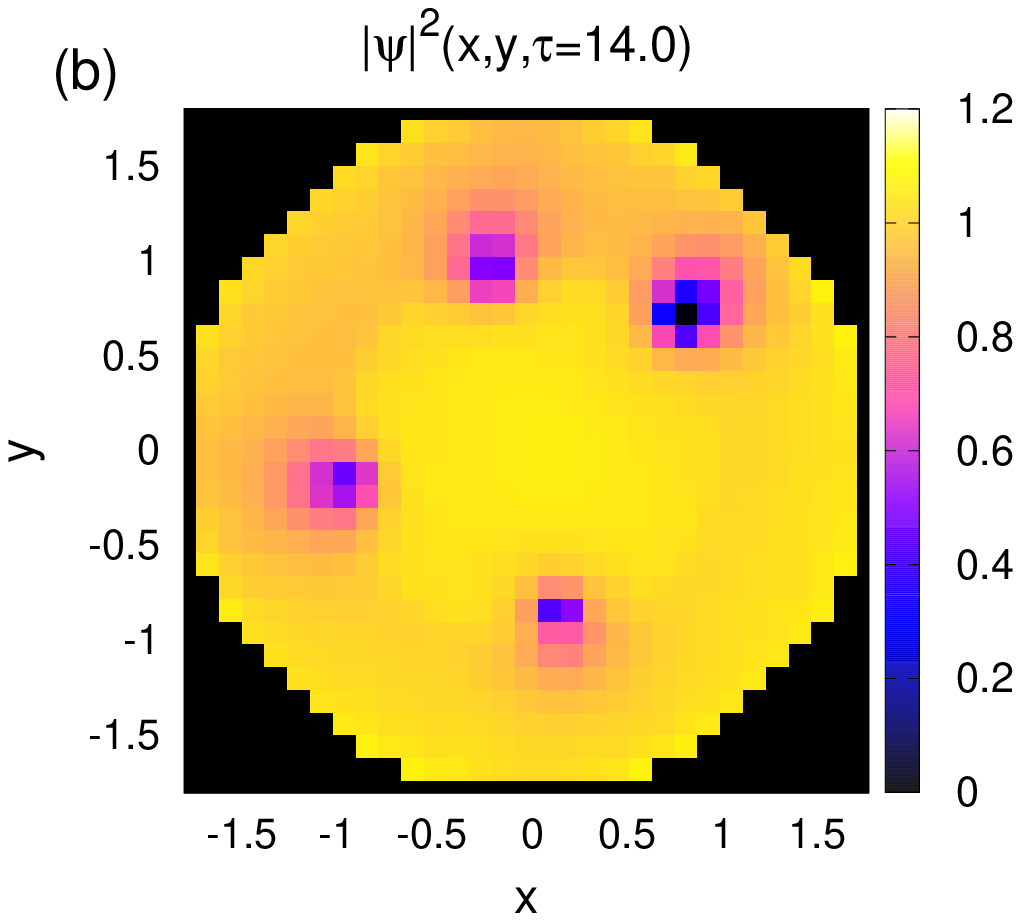, width=42mm}\\
\epsfig{file=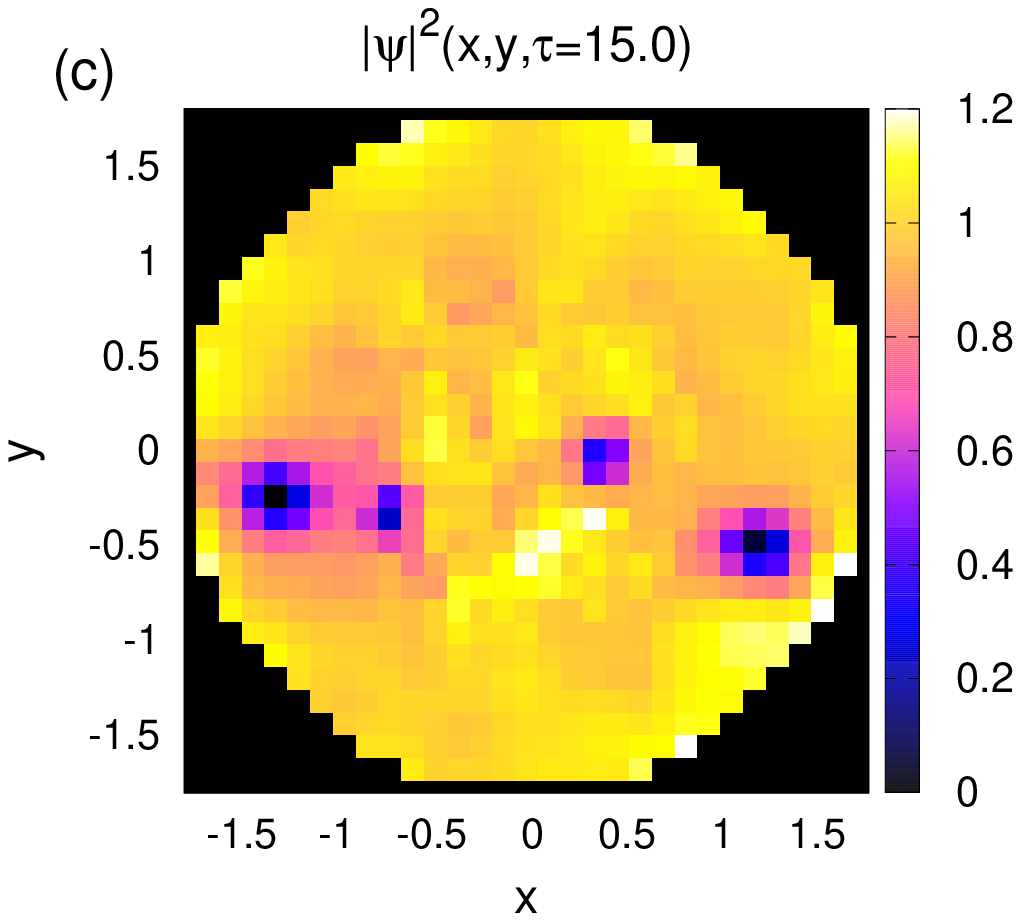, width=42mm}
\epsfig{file=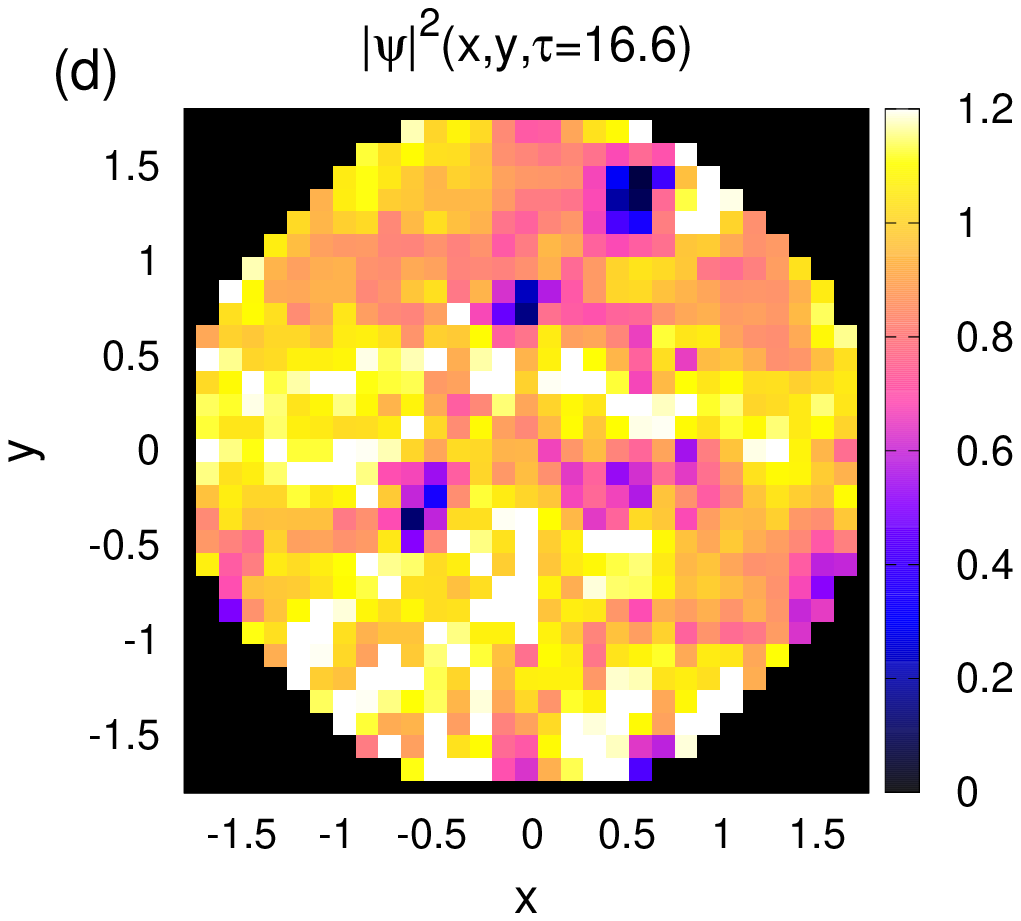, width=42mm}
\end{center}
\caption{An example of evolution of four vortices in DNLSE.}
\label{N4} 
\end{figure}

\section{Numerical results}

Eq.(\ref{psi_eq_tau}) has been numerically simulated using a 4-th order Runge-Kutta scheme 
for time stepping. Function $F(r)$ was taken in the above described simple form, with $B=3.0$.
That corresponds to using just two kinds of coupling capacitors $C_{n,n'}$. 
Thus we have a compact planar structure with a finite number of interacting 
degrees of freedom.

We present numerical results for $N=2$, $N=3$ and $N=4$ vortices 
(Figs. \ref{N2}, \ref{N3}, and \ref{N4}, respectively, where each vortex is seen 
as a density depletion). The parameters in these numerical experiments were: 
$h=0.12$, $\xi=0.05$, and $\delta=0.04$. As initial states, 
we took non-symmetric vortex configurations corresponding to numerically found 
local minima of Hamiltonian (\ref{H_xi_delta}).

The most regular dynamics was observed for $N=2$, perhaps because the simplified 
continuous counterpart (\ref{H_v}) is an integrable system in the case of two vortices 
(besides the Hamiltonian, the angular momentum is conserved). 
After initial quasi-static period of evolution [Fig.\ref{N2}(a)], there was stage of 
oscillatory motion without orbiting [Fig.\ref{N2}(b)]. Then, it was orbiting in 
anticlockwise azimuthal direction, with gradually widening cores 
[Figs.\ref{N2}(c) and \ref{N2}(d)]. Finally,  wide vortices comparable to 
the whole system size were transformed to a wave structure propagating mainly clockwise
[Figs.\ref{N2}(e) and \ref{N2}(f)]. The last stage was practically in linear regime 
because the effective nonlinear coefficient $\exp(-2\delta\tau)/\xi^2$  was very small
at $\tau\gtrsim 100$.

Vortex clusters with $3\leq N\leq 5$ passed similar initial two stages in their evolution,
but the subsequent dynamics was different. The first stage was again a stable, 
nearly static configuration, when vortex centers were motionless while their cores 
gradually broadened according to Eq.(\ref{tilde_xi}) [Figs.\ref{N3}(a) and \ref{N4}(a)]. 
The second stage was oscillation of vortices around their previous positions 
[Figs.\ref{N3}(b) and \ref{N4}(b)]. At the third stage, vortices lose stability and begin 
to move in a complicated manner, typically one or two  of them at fast ``external'' orbits 
[Figs.\ref{N3}(c) and \ref{N4}(c)]. At the fourth stage, the external vortices quit 
the lattice producing strong short-scale non-vortical oscillations in it 
[Figs.\ref{N3}(d) and \ref{N4}(d)]. During a further evolution, some of the remaining 
vortices go to external orbits and leave the lattice in a similar manner, until one or two 
are present on a highly disturbed background (not shown). 

Static initial configurations with $N\geq 6$ were not found with the given parameters.
However, cases $N=6$ and $N=6+1$ (hexagon plus central vortex) were successfully simulated 
with $h=0.04$, $\xi=0.025$, and $\delta=0.02$ (not shown). 
It should be noted that for this case the quality factor should be extremely high, since
$\gamma/g A_0^2=\delta\xi^2\sim 10^{-5}$, while $g A_0^2\sim 0.1$.
The dynamics was qualitatively  similar to that described above. It is interesting to note 
that in the last case,  the central vortex lost stability first and quickly passed 
to external orbit, crossing the system boundary soon after that.

\begin{figure}
\begin{center}
\epsfig{file=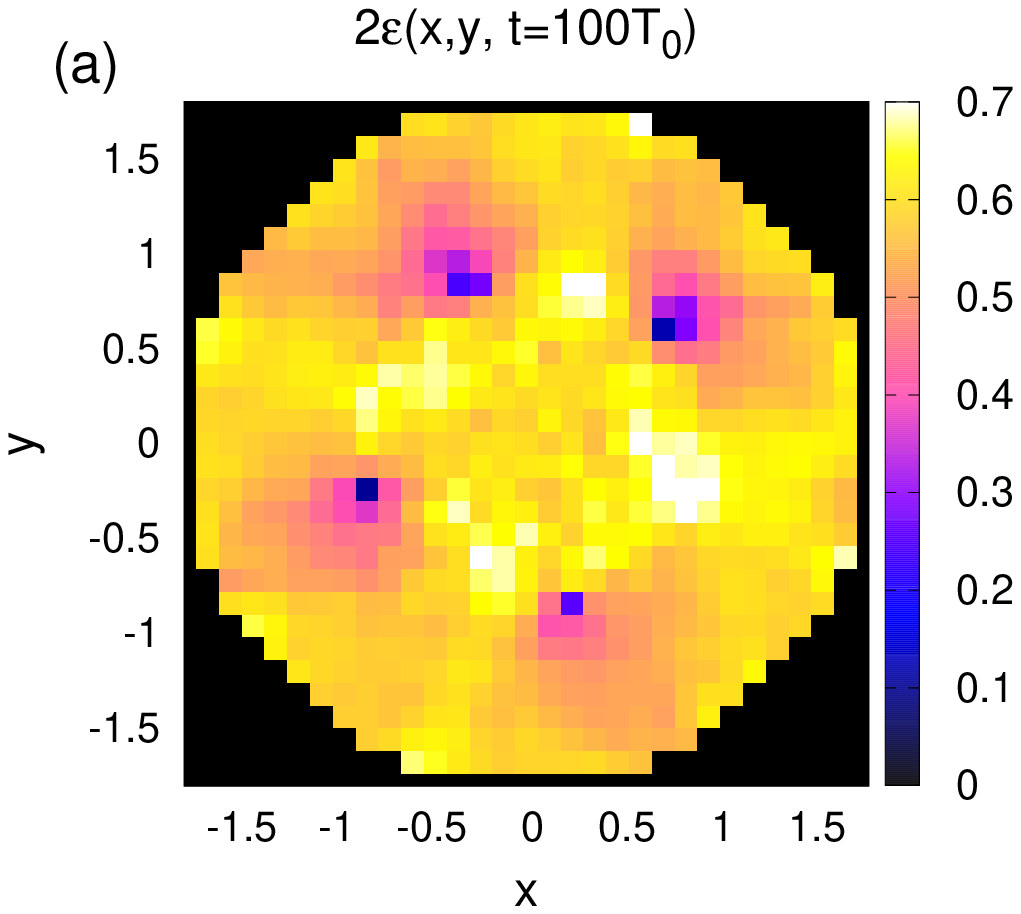, width=42mm}
\epsfig{file=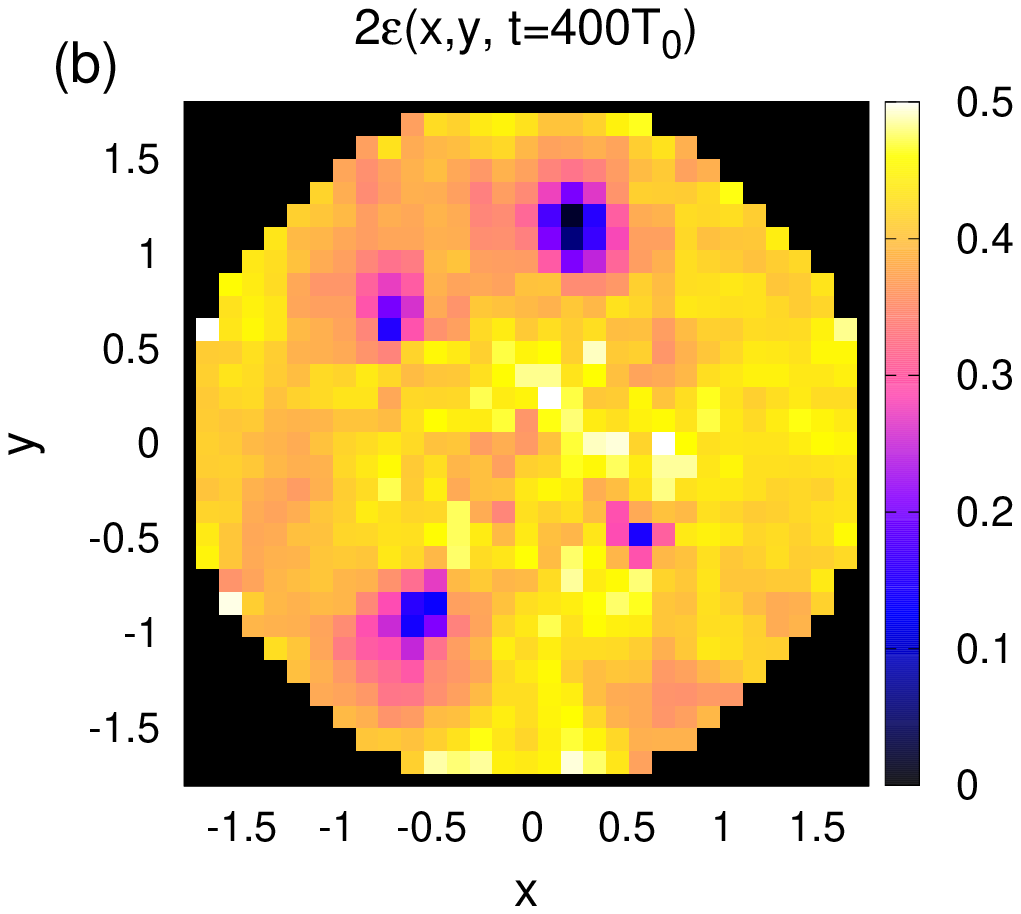, width=42mm}\\
\epsfig{file=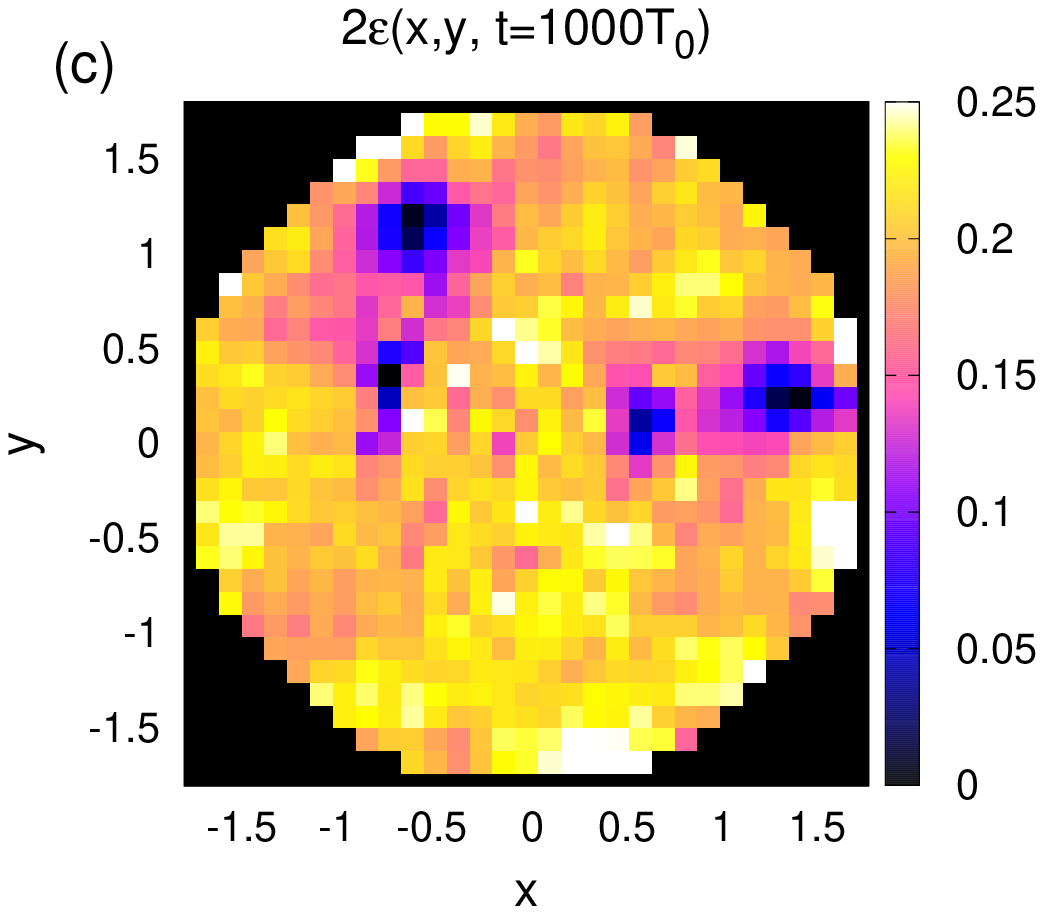, width=42mm}
\epsfig{file=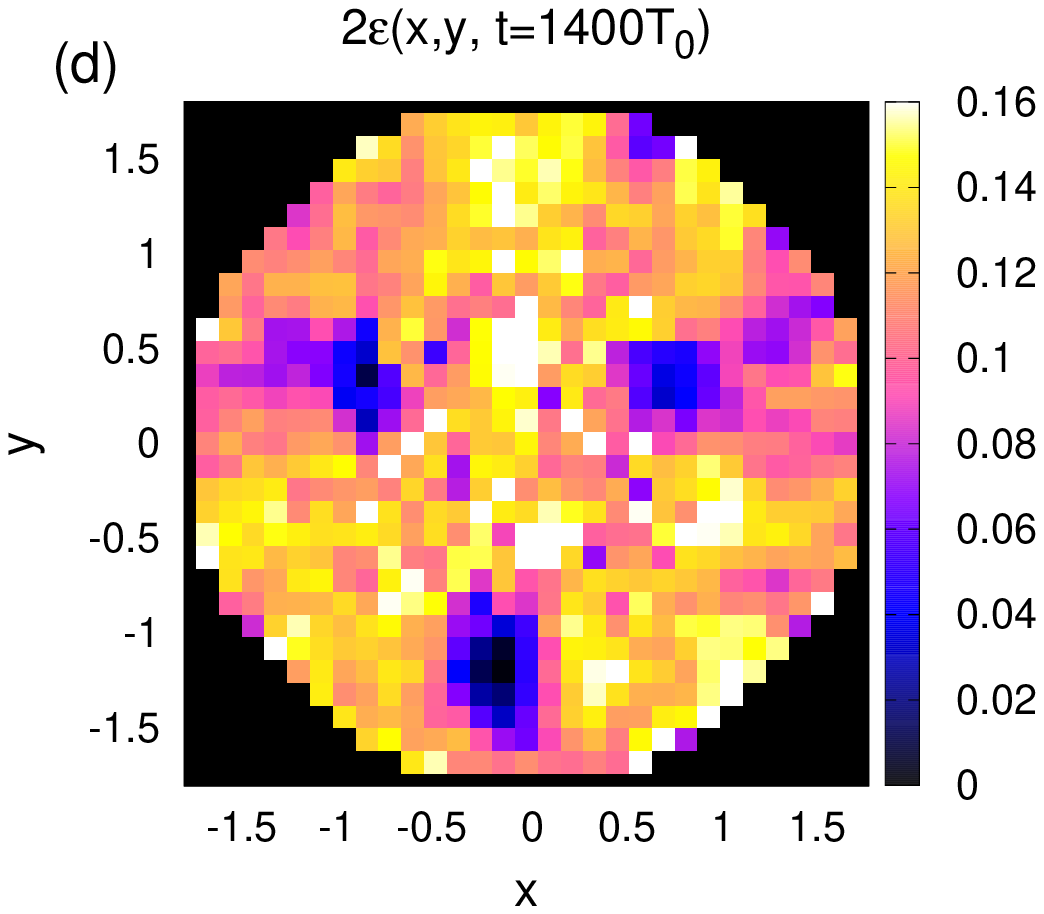, width=42mm}
\end{center}
\caption{An example of evolution of four vortices in the basic electric model.}
\label{VI_N4} 
\end{figure}
\begin{figure}
\begin{center}
\epsfig{file=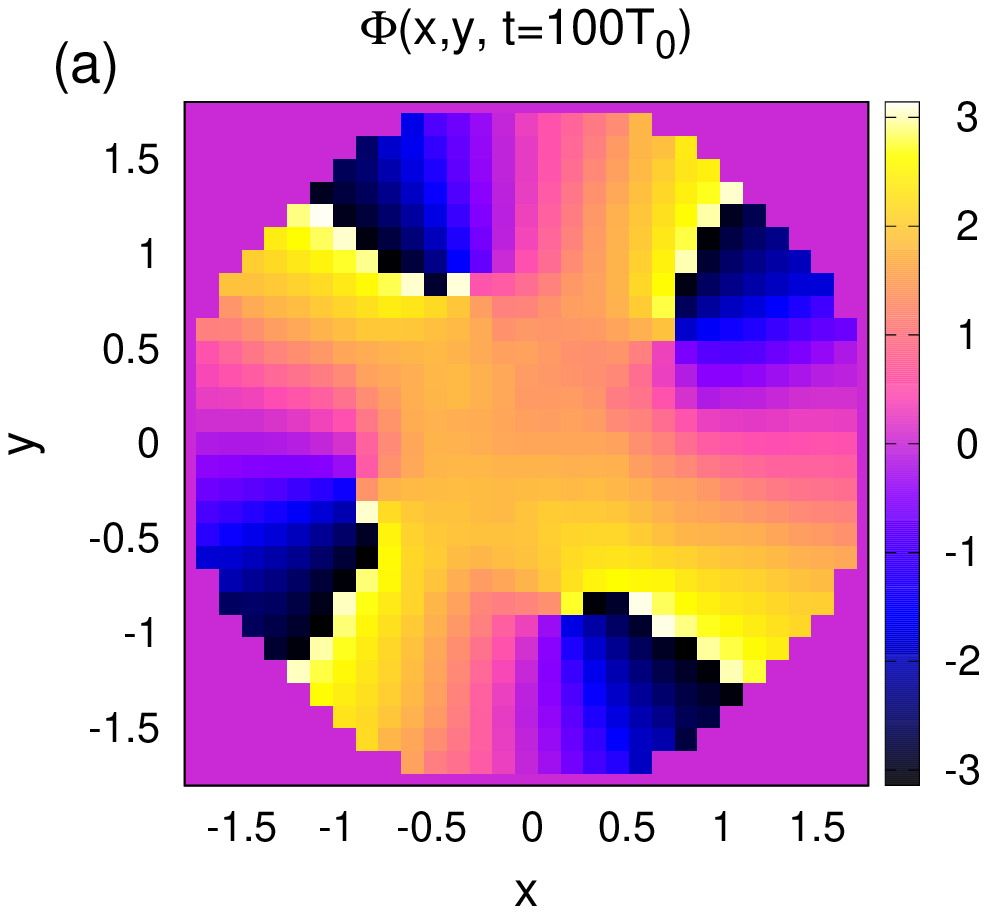, width=42mm}
\epsfig{file=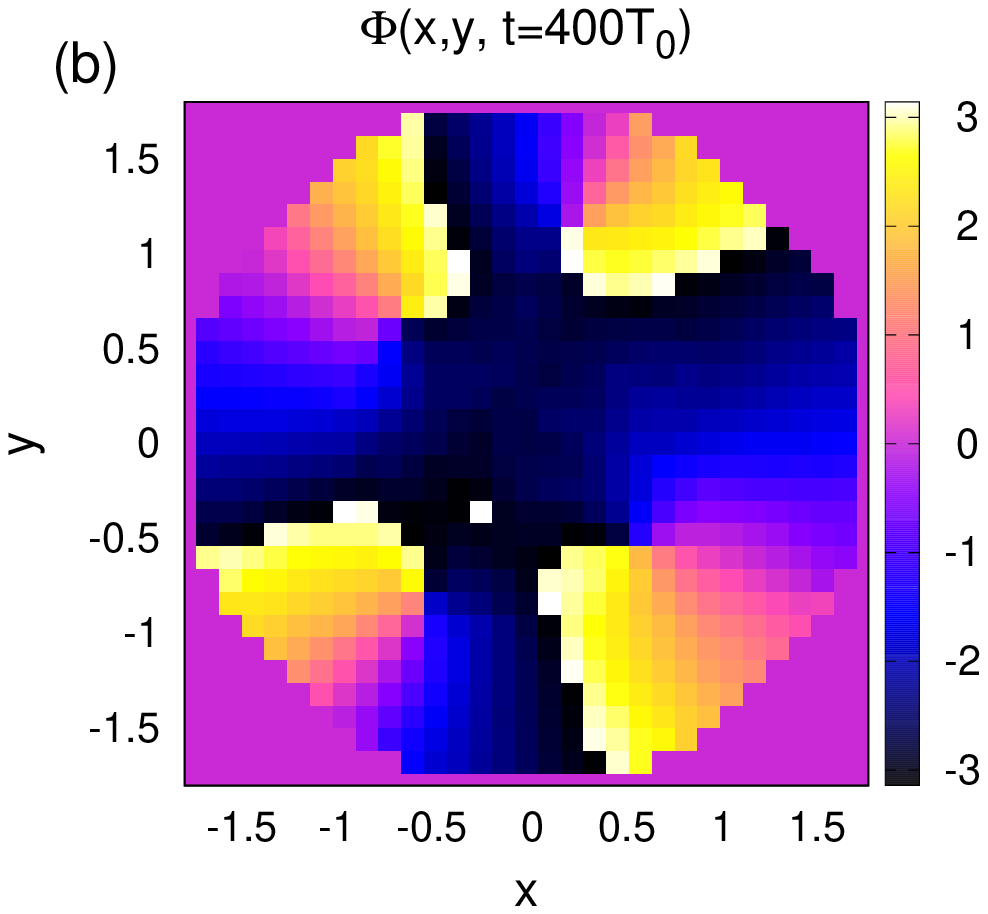, width=42mm}\\
\epsfig{file=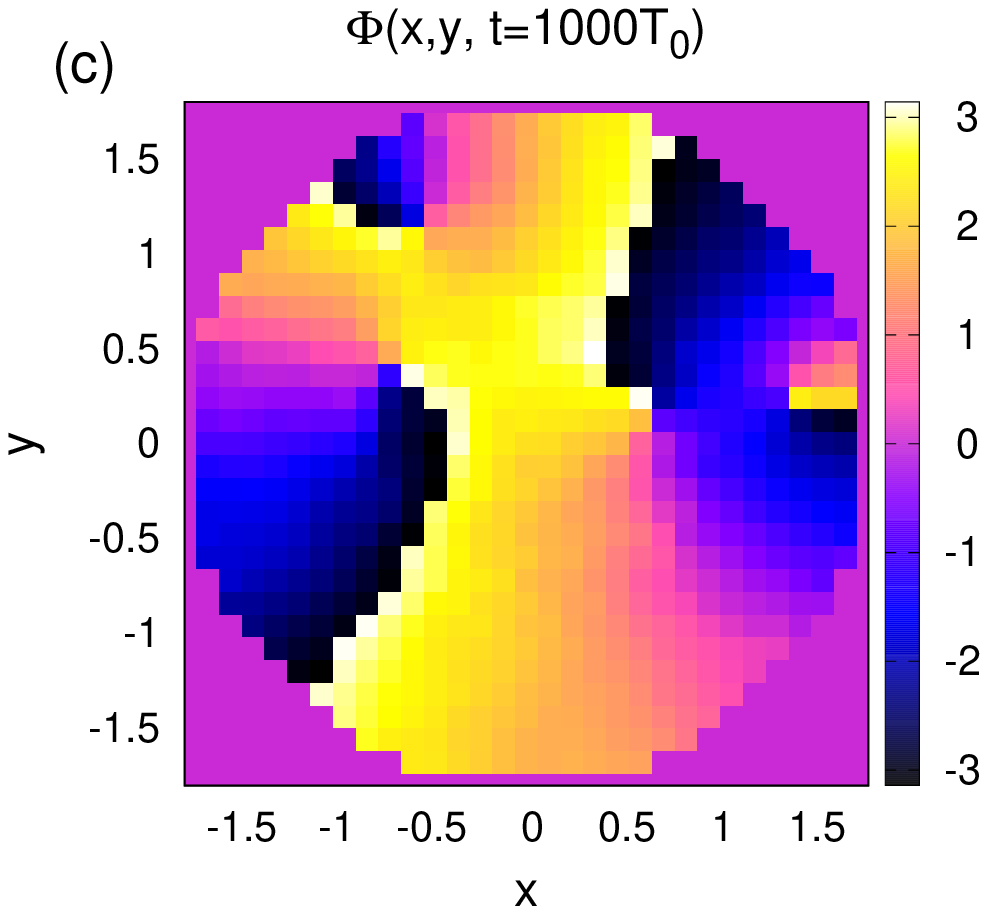, width=42mm}
\epsfig{file=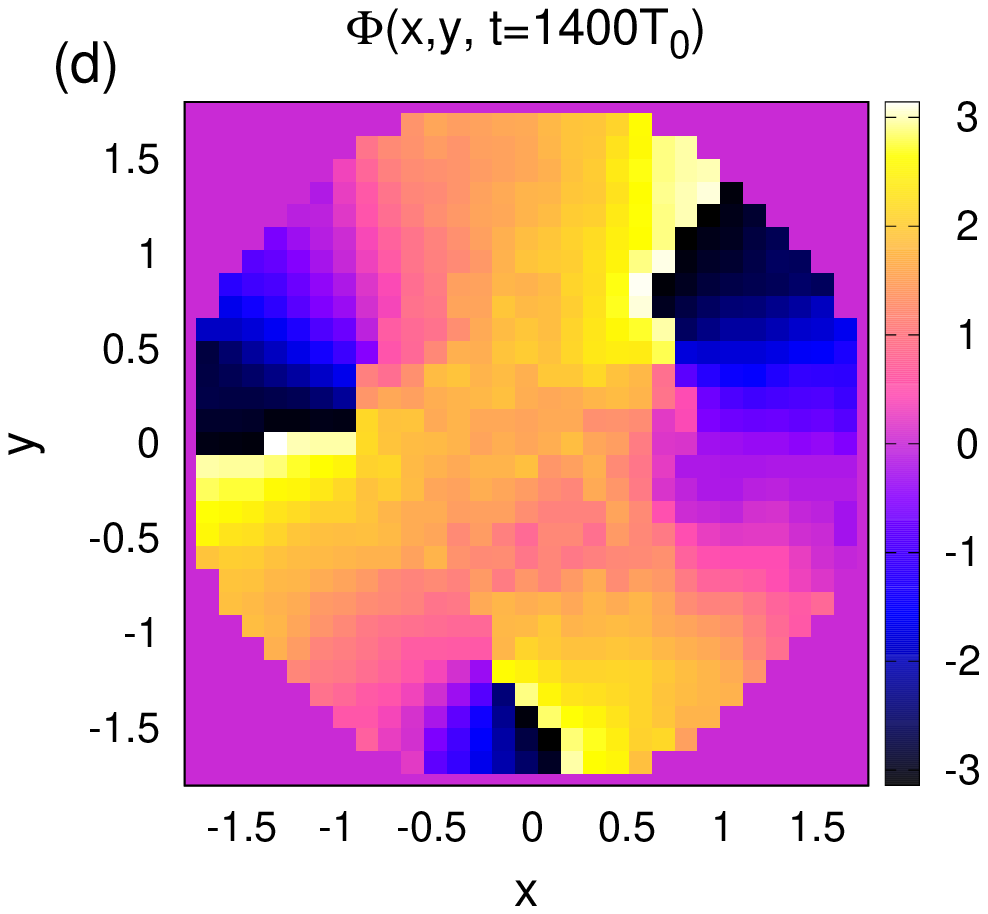, width=42mm}
\end{center}
\caption{Four vortices in the electric model: phases corresponding to Fig.\ref{VI_N4}.
The presence of vortices is clearly seen.}
\label{VI_N4-phase} 
\end{figure}
\begin{figure}
\begin{center}
\epsfig{file=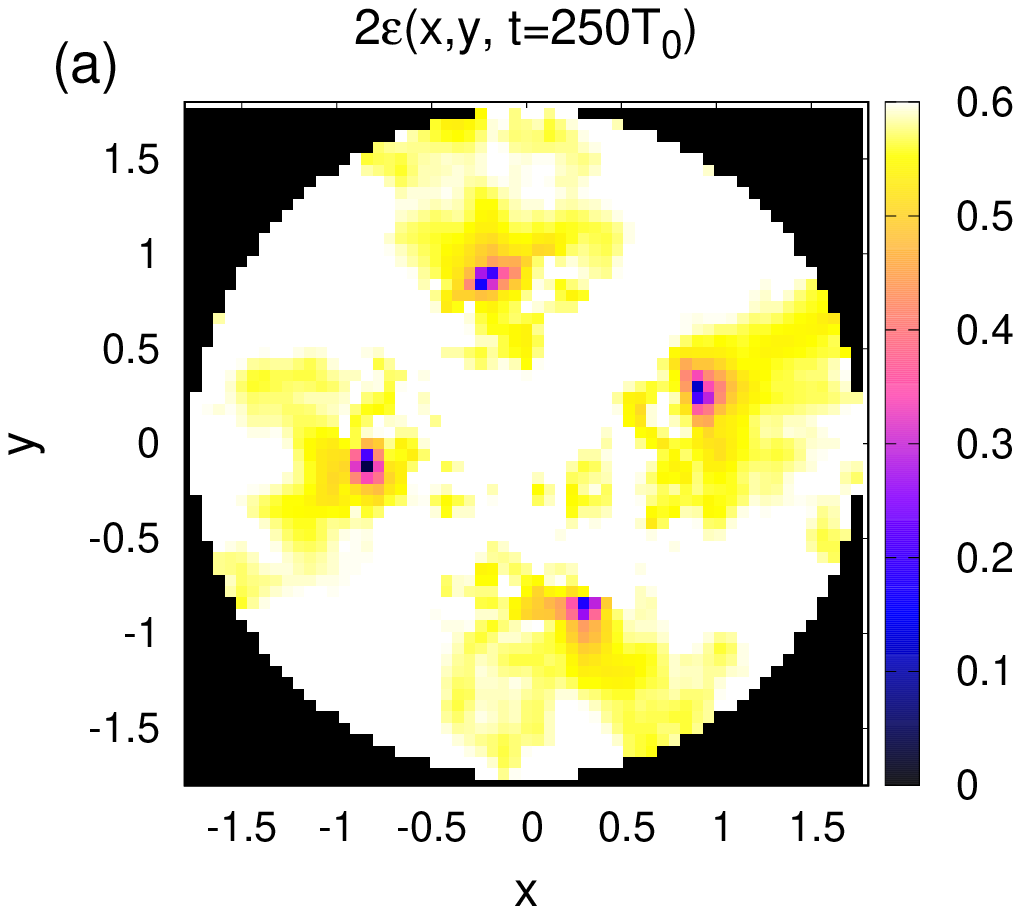, width=42mm}
\epsfig{file=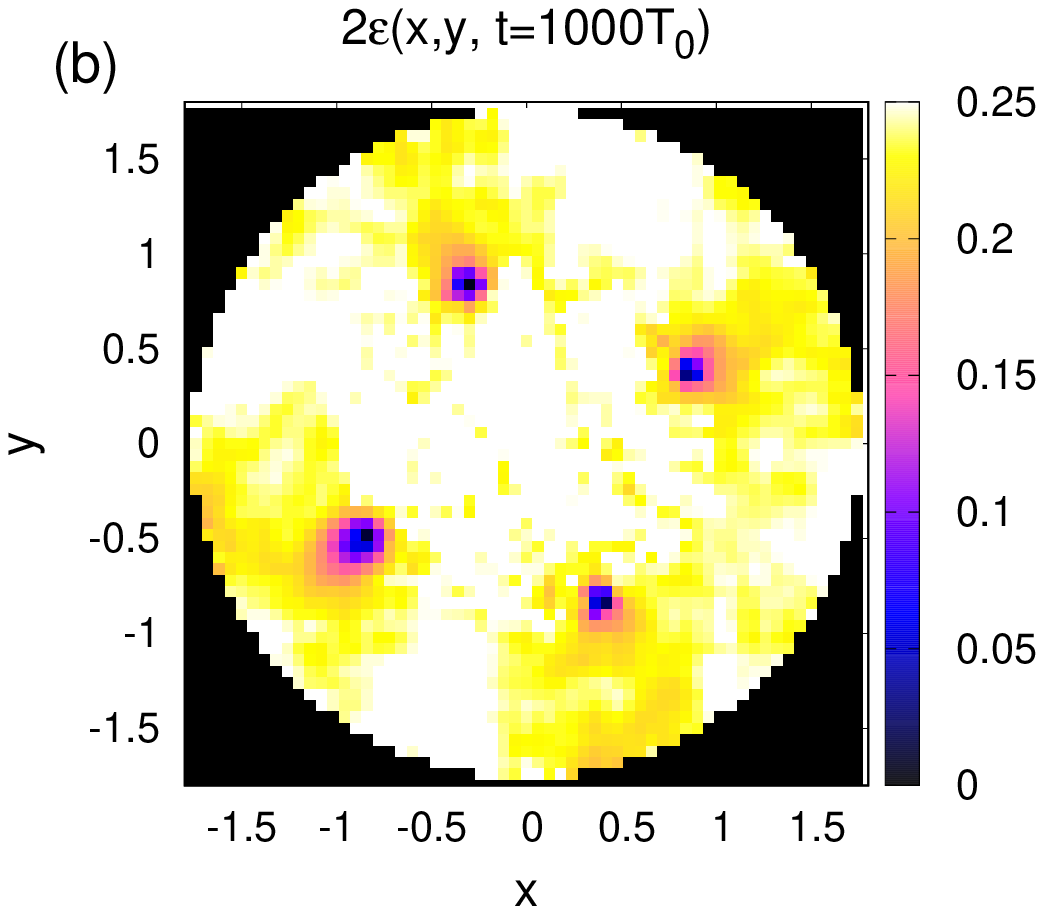, width=42mm}\\
\epsfig{file=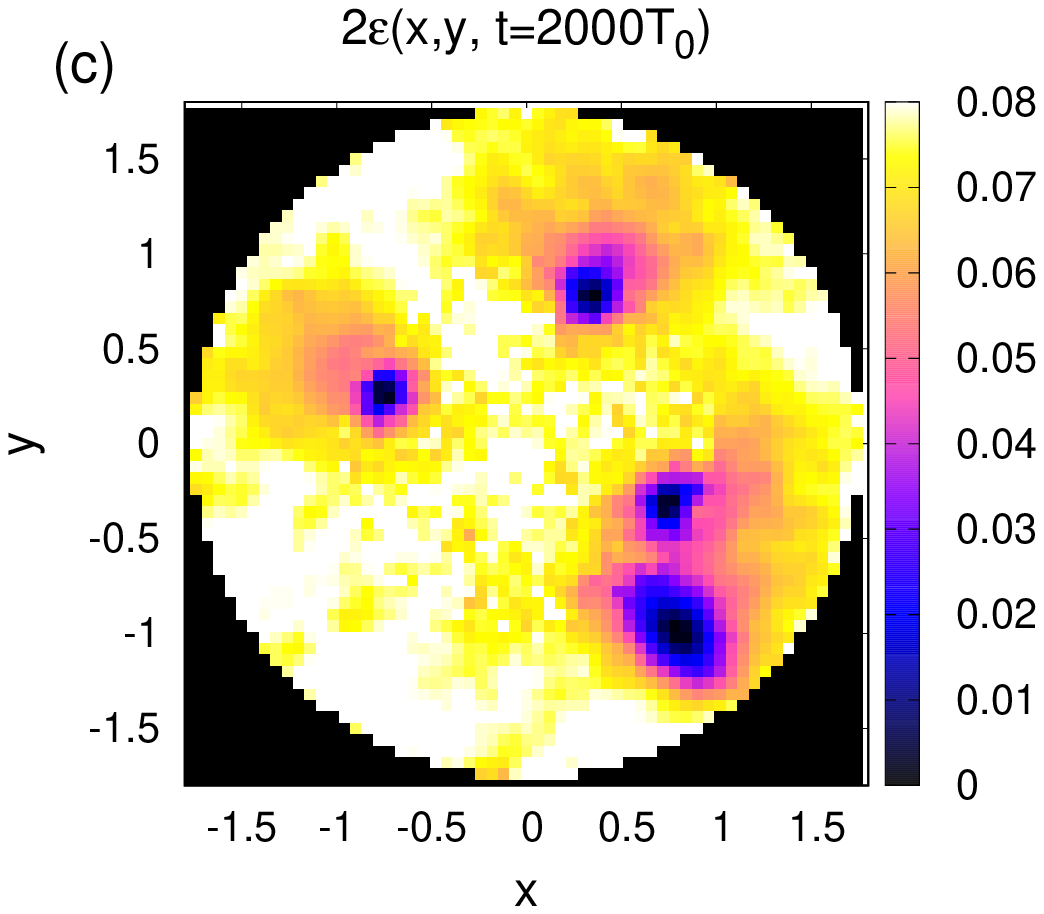, width=42mm}
\epsfig{file=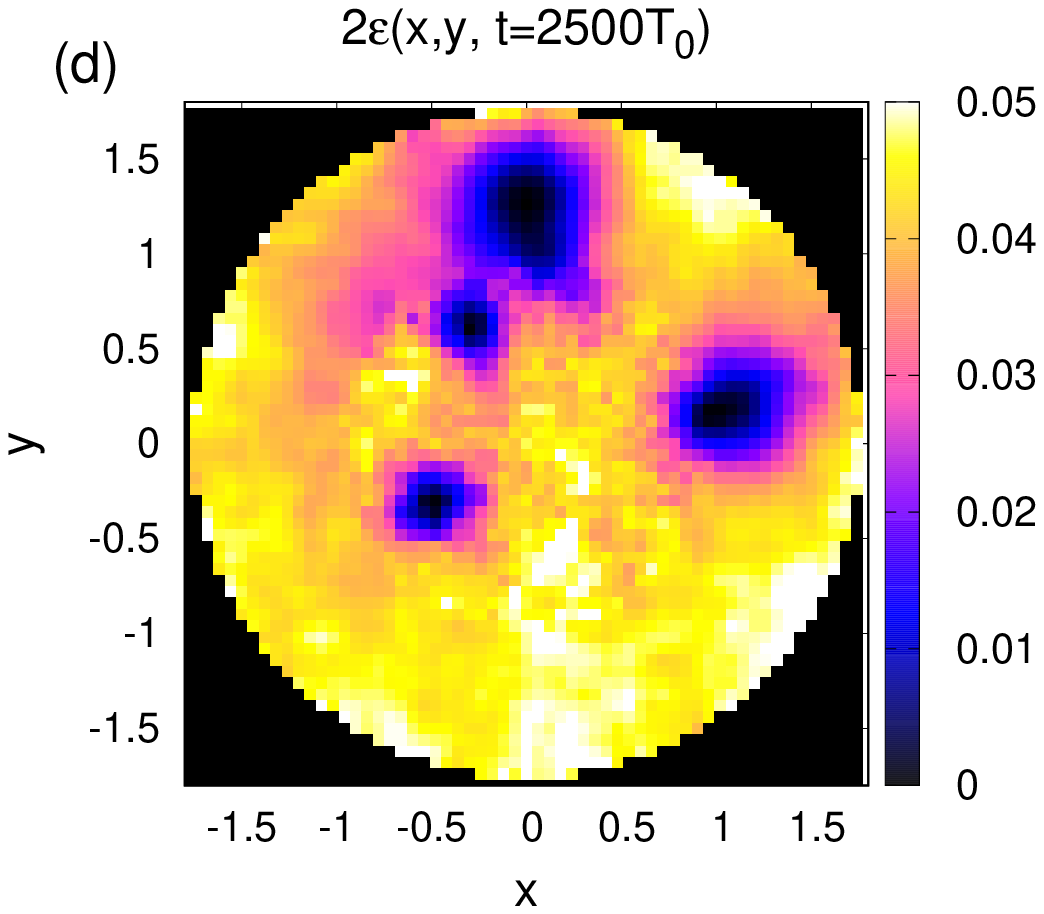, width=42mm}
\end{center}
\caption{An example of evolution of four vortices in the electric model with smaller $h=0.06$.}
\label{VI_N4new} 
\end{figure}

Of course, the above results were obtained within DNLSE under many simplifying assumptions, 
and therefore they cannot be completely convincing. In order to get a more direct evidence
of vortex existence and behavior in fully nonlinear regime, the original system of circuit
equations (\ref{current})-(\ref{voltage}) has been numerically simulated using
expression (\ref{C_V}) with parameters $\eta=0$, $\nu=2$, $\mu=0.5$ (and $C_0=1$, $V_*=1$).
Two numerical experiments are presented below. In the first one 
(see Figs.\ref{VI_N4}-\ref{VI_N4-phase}), the remaining
dimensionless parameters were $\bar c=0.02$, $h=0.12$, $L=1$, $R_L=10^{-4}$,
$R_C=10^{4}$. At $t=0$, the partial energies of oscillators were corresponding to
$I_n^2/2+W(V_n)=0.32$ (excluding vortex cores), while their ``phases'' 
$\Phi_n=\arctan(I_n/V_n)$ were the same as the initial phases for DNLSE simulation. 
Therefore $|gA_0^2|\approx (5/24)\cdot 0.32\approx 0.067$ in these numerical experiments.
Initial a.c. voltages were in the range $-0.6\lesssim V_n\lesssim 1.0$.

To resolve Eqs.(\ref{current}) with respect to $\dot V_n$, a simple iterative
scheme was developed,
\begin{eqnarray}
D_n^{(j+1)}&=&D_n^{(j)}-0.2\Big[C(V_n)D_n^{(j)}\nonumber\\
&+&\sum_{n'} C_{n,n'}(D_n^{(j)} -D_{n'}^{(j)})
+\frac{V_n}{R_C}- I_n\Big],
\end{eqnarray}
with $D_n^{(0)}=(I_n-V_n/R_C)/C(V_n)$. The result of 60-th iteration 
$\dot V_n\approx D_n^{(60)}$ was then used in a Runge-Kutta 4-th order time stepping.
The convergence of this scheme was ensured
by positive definiteness of the corresponding quadratic form, and  by 
choosing the coefficient $0.2$ sufficiently small to have $|1-0.2\cdot C_{max}|<1$ 
(where $C_{max}=C(V_{min})$, and $V_{min}\approx -0.57$ is the negative root of equation
$W(V)=0.32$). 

In Fig.\ref{VI_N4}, the evolution of quantities $2\varepsilon_n=I_n^2+2W(V_n)$ 
is shown for the case of four vortices, while in Fig.\ref{VI_N4-phase} we see 
the corresponding phases. In particular, Fig.\ref{VI_N4-phase} indicates unambiguously
that we deal with vortices, not simply with some amplitude depressions.
Qualitatively, the system passed the same stages as in the simplified DNLSE model. 
However, since here the initial phase distribution was not
appropriately adjusted to strong nonlinearity, the first (trapping) stage was not so 
long as in experiment shown in Fig.\ref{N4}.

\begin{figure}
\begin{center}
\epsfig{file=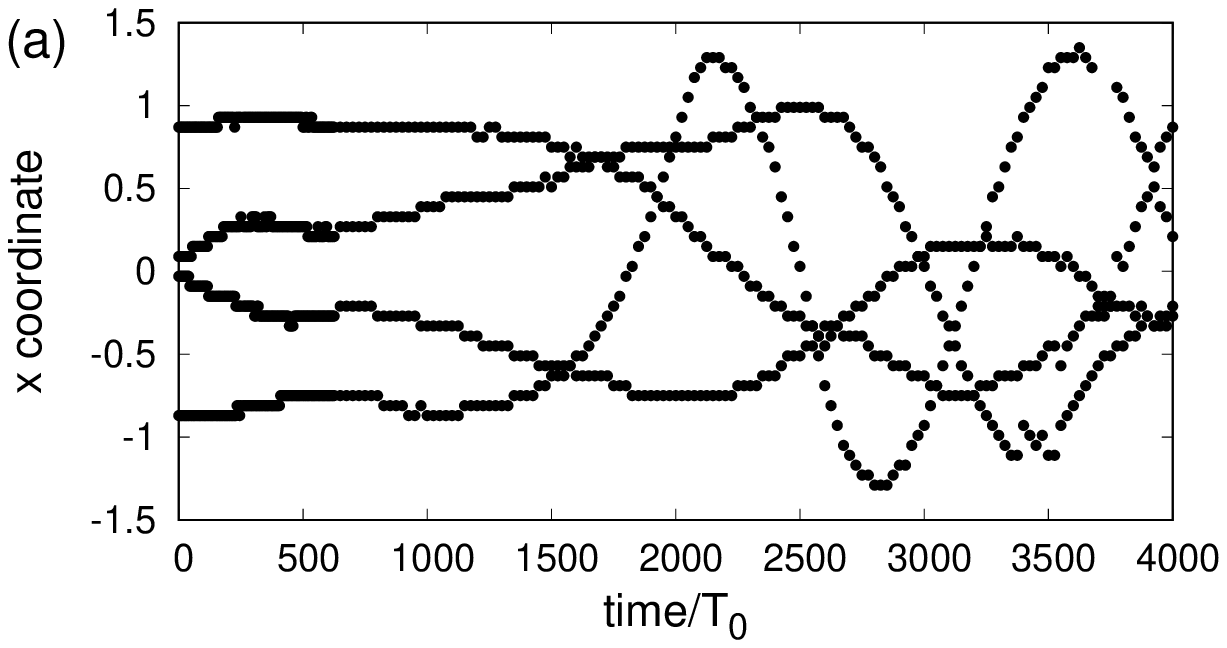, width=80mm}\\
\epsfig{file=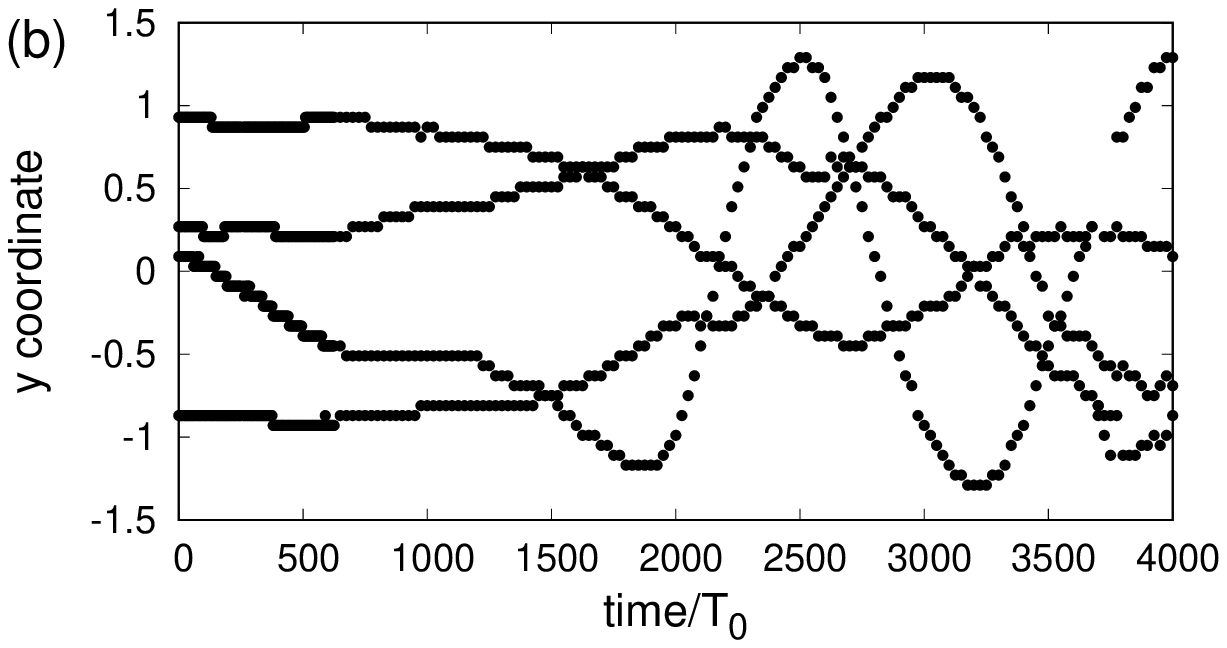, width=80mm}
\end{center}
\caption{Time-dependence of $x$ and $y$ coordinates of four vortices, corresponding to
simulation with $h=0.06$. Conventionally, vortices are located at the centers of those 
$h\times h$ squares, where the sum of phase increments along the sides is $2\pi$.}
\label{VI_N4new-xy} 
\end{figure}

Finally, in Figs.\ref{VI_N4new}-\ref{VI_N4new-xy} we present results for a smaller $h=0.06$ 
and for larger initial energies $I_n^2+2W(V_n)=0.81$. In this simulation $\bar c=0.04$. 
In general, vortices look more smooth here. As Figs.\ref{VI_N4new}a-b demonstrate, and 
Fig.\ref{VI_N4new-xy} confirms, the cluster was almost static till $t\sim 1000 T_0$.
After that time the configuration was deformed by appeared instability, 
and the vortices started intense motion. Unlike the case $h=0.12$, here no vortex exited 
the disc till the very end of simulation.

\section{Summary}

To summarize, in this work a general scheme of an electric network has been suggested 
which can be approximately described by a weakly dissipative defocusing discrete nonlinear 
Schr\"odinger equation of special kind, where coupling terms are not translationally 
invariant but spatially uniform background solutions exist. 
Discrete vortices in such systems have been analyzed and then numerically simulated.
Simulations have demonstrated qualitatively similar results
within DNLSE and within the original circuit equations.

Of crucial importance is the quality factor of oscillator circuits. Numerical experiments 
have shown that nontrivial behavior of vortices is observable with $Q\gtrsim 10^4-10^5$.
In practice such values could be achieved at sufficiently low temperatures when conductivity
of metals as well as resistivity of dielectrics are both substantially higher than they are
at the room temperature.

The study above is apparently far from being exhaustive.
This system seems deserving further thorough investigation, especially in its highly
nonlinear regimes, and under external driving (driving signals can be easily introduced
into electric network, resulting in many resonance phenomena, perhaps similar in some 
sense to those reported in Ref.\cite{RVSDK2012}). 
The author also hopes that experimentalists will be interested 
in conducting laboratory experiments inspired by the present theory.

\end{document}